\pdfoutput=1
\documentclass[12pt]{article}

\usepackage{graphicx,psfrag,epsf,color}
\usepackage{amsmath,amssymb,amsfonts}
\usepackage{array}
\usepackage{cite}
\usepackage{multirow}
\usepackage{rotating}
\usepackage{slashed}
\usepackage{todonotes}
\usepackage{bm}
\numberwithin{equation}{section}
\newcounter{MBQ}

\def\slash#1{#1 \hskip-0.45em /}

\newcommand{\be}{\begin{equation}}
\newcommand{\ee}{\end{equation}}
\newcommand{\bea}{\begin{eqnarray}}
\newcommand{\eea}{\end{eqnarray}}
\newcommand{\bi}{\begin{itemize}}
\newcommand{\ei}{\end{itemize}}
\newcommand{\ben}{\begin{enumerate}}
\newcommand{\een}{\end{enumerate}}
\newcommand{\bt}{\begin{tabular}}
\newcommand{\et}{\end{tabular}}

\newcommand{\Eres}{E^\gamma_{\rm res}}
\newcommand{\mchi}{m_\chi}

\newcommand{\schr}{Schr\"{o}dinger}

\newcommand{\tev}{{\rm TeV}}

\def\lmuh{l_{\mu_h}}

\def\lmus{l_{\mu_s}}

\newcommand{\nn}{\nonumber}

\begin{document}
\allowdisplaybreaks

\begin{titlepage}

\begin{flushright}
{\small
TUM-HEP-1240/19\\
arXiv:1912.02034\\[0.0cm]
December 16, 2020
}
\end{flushright}

\vskip1cm
\begin{center}
{\Large \bf Precise yield of high-energy photons from \\[0.1cm] 
Higgsino dark matter annihilation}\\[0.2cm]
\end{center}

\vspace{0.5cm}
\begin{center}
{\sc M.~Beneke$^{a}$, C.~Hasner$^{a}$, K.~Urban$^{a}$, } \\ 
and  {\sc M.~Vollmann$^{a}$}\\[6mm]
{\it ${}^a$Physik Department T31,\\
James-Franck-Stra\ss e~1, 
Technische Universit\"at M\"unchen,\\
D--85748 Garching, Germany}
\\[0.3cm]
\end{center}

\vspace{0.6cm}
\begin{abstract}
\vskip0.2cm\noindent
The impact of electroweak Sudakov logarithms on the endpoint of the photon spectrum for wino dark matter annihilation was studied intensively over the last several years. In this work, we extend these results to Higgsino dark matter~$\chi_1^0$. We achieve NLL' resummation accuracy for narrow and intermediate spectral energy resolutions, of order $m_W^2 / \mchi$ and $m_W$, respectively. This is the most accurate prediction to date for the yield of high-energy $\gamma$-rays from $\chi_1^0 \chi_1^0 \to \gamma + X$ annihilation for the energy resolutions realized by current and next-generation telescopes. We also discuss for the first time the effect of power corrections in $m_W/\mchi$ in this context and argue why they are not sizeable.
\end{abstract}
\end{titlepage}

\section{Introduction}
\label{sec:introduction}
In recent years much attention has been devoted to understanding the nature of the dark matter (DM) in the Universe. 
This type of matter is not accounted for in the Standard Model (SM) of particle physics and its existence is supported by compelling evidence from Galactic \cite{Rubin:1980zd} to Cosmological scales \cite{Ade:2015xua}.

Due to their accidental relationship with the electroweak scale, DM candidates known as weakly interacting massive particles (WIMPs) \cite{Jungman:1995df} have received most of the attention. 
Negative results from direct and indirect searches \cite{Arcadi:2017kky,Roszkowski:2017nbc} of these WIMPs especially in the electroweak (EW) mass scale range suggest that more attention should be paid to less explored avenues such as the multi-TeV WIMP realizations. 

A benchmark example for such heavy candidates is the Higgsino-like neutralino in supersymmetric extensions of the SM \cite{ArkaniHamed:2006mb} and the limit of the pure Higgsino DM model, which extends the SM by a single fermionic SU(2) doublet. Its mass has to be approximately 1~TeV if its cosmological abundance is to be explained by a freeze-out process. As a consequence of this large mass and the fact that in this scenario the Higgsino is then the lightest supersymmetric particle (LSP), the Higgsino DM model does not quite address the naturalness problem. The attractiveness of the model is its simplicity. However, direct searches for TeV-mass Higgsinos with the LHC are almost impossible and their scattering cross section with nucleons $\sigma_{\rm SI}\sim 10^{-48}$~cm$^2$ \cite{Hill:2014yxa,Hisano:2015rsa} is below the so-called neutrino floor in direct DM detection experiments. 

Nevertheless, DM models of the Higgsino type can be discovered indirectly if a line signature in the gamma-ray spectrum from, for example, the innermost region of our Galaxy, is observed. The quasi-monochromaticity of this part of the spectrum is a consequence of the kinematics of pair annihilation of non-relativistic WIMPs---a key prediction of many WIMP models including the Higgsino. This signal is particularly interesting because disentangling it from the uncertain astrophysical foregrounds is easier than for other type of spectra. Moreover, there is no known astrophysical mechanism that features all the aforementioned properties, rendering the observation of such spectral-line signals a smoking-gun discovery of annihilating DM.

Searches for this type of signature have already been performed by existing gamma-ray telescopes such as Fermi-LAT \cite{Ackermann:2015lka}, H.E.S.S. \cite{Abdallah:2018qtu} and MAGIC \cite{Aleksic:2013xea}. 
Particularly promising is the next-generation Cherenkov Telescope Array (CTA) \cite{Acharya:2017ttl}, expected to improve the existing limits on spectral lines by one order of magnitude for TeV-scale gamma-ray energies. In these searches, though, it is assumed that the endpoint gamma-ray spectrum from DM annihilation is dominated by the fully-exclusive $\chi_1^0\chi_1^0\to\gamma\gamma$ process. Under this assumption, the endpoint signal would be a perfectly monochromatic signal at the gamma-ray energy $E_\gamma=m_\chi$ as required by kinematics. 

Even though the monochromatic approximation might be a reasonable assumption for some (light) WIMP models, it is not for generic models. The unavoidable finite energy resolution and the fact that only a single photon is observed at a time, imply that the proper observable is the flux that originates from the $\chi_1^0\chi_1^0\to\gamma+X$ process, where $X$ is any type of (undetected) primary and secondary radiation that can be emitted in association with the observed gamma ray, and where the energy of the detected photon is close 
to the maximal value $m_\chi$ within the energy resolution.

Theoretical predictions of spectral-line signals and the semi-inclusive photon spectrum near maximal photon energy for heavy WIMPs are not as straightforward as it might appear, since the perturbative expansion in the EW coupling constants breaks down. The effects giving rise to this are twofold. First, one has to take into account the so-called Sommerfeld effect, which is generated by the electroweak Yukawa force acting on the DM particles prior to their annihilation \cite{Hisano:2003ec,Hisano:2004ds,ArkaniHamed:2008qn,Beneke:2014hja}. 
Secondly, for heavy DM annihilation into energetic particles, electroweak Sudakov (double) logarithms $\mathcal{O}((\alpha_2 \ln^2(\mchi/m_W)))$ are large and need to be resummed to all orders in the coupling constant\cite{Baumgart:2014vma,Ovanesyan:2014fwa,Bauer:2014ula,Ovanesyan:2016vkk,Baumgart:2017nsr,Beneke:2018ssm,Beneke:2019vhz}. The treatment of the Sommerfeld effect for Higgsino DM is well known. In this work, we focus on the resummation of large logarithmic corrections in the annihilation process $\chi_1^0 \chi_1^0 \rightarrow \gamma + X$ for Higgsino DM, using soft collinear effective theory (SCET).

In recent papers we \cite{Beneke:2018ssm,Beneke:2019vhz} and others\cite{Baumgart:2014vma,Ovanesyan:2014fwa,Ovanesyan:2016vkk,Bauer:2014ula,Baumgart:2017nsr} discussed these effects for the  pure electroweak Majorana triplet model (wino-like neutralino)\cite{Cirelli:2009uv}. In those papers we established factorization formulas and obtained precise predictions for the photon yield for two different telescope energy-resolution regimes by using SCET methods \cite{Bauer:2000yr,Bauer:2001yt,Beneke:2002ph,Beneke:2002ni}. 
We also provided the necessary functions for the evaluation of the factorization formulas at the next-to-leading logarithmic prime (NLL') accuracy. 

Here we extend these results to the case of the Higgsino DM, which is slightly more involved, due to the non-zero hypercharge of the Higgsino. As already mentioned, if the thermally produced Higgsino is to make up all the DM in the Universe, its mass should be approximately 1 TeV. This mass is indeed much larger than the masses of the EW gauge bosons and resummation of large Sudakov logs is necessary. Nevertheless, it is three times smaller than the mass of the thermally produced wino DM. As a consequence, $m_W/\mchi$ mass corrections that are systematically neglected in the SCET leading-power resummations developed up to now might be as large as (or even larger than) the percent-level accuracy of NLL' resummation.
We thus include for the first time a quantitative discussion of the next-to-leading power mass correction for Higgsino DM (with direct applicability for wino DM as well). 

The resummation of large Sudakov logarithms is achieved by breaking down the annihilation cross section into a few calculable factors representing the different scales and momentum configurations that build up the large logarithms. In order to prove such factorization formulas, some assumptions on the resolution $\Eres$ of the photon energy measurement are necessary. In \cite{Beneke:2019vhz} we distinguished three parametrically different resolution regimes:\footnote{The theoretical 
framework for the ``narrow'' regime also applies to 
$\Eres \ll m_W^2/\mchi$ including the classic $\gamma\gamma$ line 
with $\Eres=0$.}
\begin{align}
\text{narrow}: \quad \Eres \sim&\, m_W^2/\mchi \nn \\
\text{intermediate}: \quad \Eres \sim&\, m_W \nn \\
\text{wide}: \quad \Eres \gg&\, m_W
\end{align}
Given the projected energy resolution of the upcoming CTA experiment \cite{Acharya:2017ttl} and the thermal Higgsino DM mass $\mchi \approx 1$~TeV, it is clear that the narrow and intermediate resolution regimes defined above are the most appropriate when computing the resummed cross section (see Figure 1 in \cite{Beneke:2019vhz}).

The article is organized as follows. In Section \ref{sec:theory} we briefly review the Higgsino model, summarize the SCET annihilation operator basis, and the factorization formula for the intermediate energy resolution. In Section \ref{sec:nlp}we consider mass corrections. In Section \ref{sec:results} we present the numerical calculation of the semi-inclusive annihilation cross section, after which we conclude in Section \ref{sec:concl}. The main text is held short and focuses on non-technical results. 
In a series of appendices we give the potential used for the computation of the Sommerfeld effect, provide details of the factorization formula in the intermediate and narrow energy resolution regime and summarize the complete NLO expressions for all relevant functions 
that enter the factorization formula.

\section{Endpoint spectrum of \boldmath$\chi_1^0\chi_1^0\to\gamma+X$}
\label{sec:theory}
The Higgsino model consists of the simple extension of the SM by 
an EW vector-like fermion SU(2) doublet with hypercharge $Y=1/2$, 
such that one component is electromagnetically neutral after EW symmetry breaking (EWSB) \cite{Cirelli:2005uq,Cirelli:2009uv}. The Lagrangian of the model is given by
\begin{equation}
\label{eq:fullth}
\mathcal{L}= \mathcal{L}_{\text{SM}} + \overline{\chi} (i \slashed{D} - m_{\chi}) \chi + \mathcal L_{\rm dim-5}\,.
\end{equation}
The SU(2)$\otimes$U(1)$_Y$ covariant derivative is defined as $D_\mu=\partial_\mu-ig_2 W_\mu^CT^C + i g_1 Y B_\mu$. 
With our conventions \eqref{eq:rotMatrix} the choice of EW charges is such that the lower component of the multiplet is neutral, that is, in terms of components 
$\chi=(\chi^+,\chi_D^0)$, where the superscript denotes the 
electric charge. The mass eigenstates after EWSB are two 
self-conjugate (Majorana) particles ($\chi_1^0$ and $\chi_2^0$) 
defined in such way that  $\chi_D^0=(\chi_1^0+i\chi_2^0)/\sqrt2$ 
and an electromagnetically charged Dirac (chargino) particle $\chi^+$.

A higher-dimension effective operator, for example
$\mathcal L_{\rm dim-5}=\frac{1}{\Lambda}(\overline{\chi}\Phi)i\gamma_5(\Phi^\dagger\chi)$ where $\Phi$ is the standard Higgs doublet,
 is necessary to provide the $\chi_2^0$ particle with a slightly (at least $\mathcal O(100)$~keV) larger mass than the $\chi_1^0$ particle.  
Otherwise, $Z$-boson mediated tree-level couplings of the Higgsinos to the light quarks would induce a large nucleon-DM cross section already ruled out by direct-detection experiments. On the other hand, 
the mass splitting $\delta m=m_{\chi^-}-m_{\chi_1^0}$ 
between the charged and neutral component of the Higgsino doublet 
is induced radiatively after EWSB. At the one-loop order 
\cite{Thomas:1998wy},  $\delta m=\alpha_{\rm em}(m_Z/2+\mathcal O(m_Z^2/\mchi))$ $\approx\,355$~MeV.

Due to the Sommerfeld effect, the annihilation cross section is very sensitive to small variations of the mass splittings. However, the resummation of the Sudakov logarithms is insensitive to these variations. For instance, we checked that the effect of NLL' resummation changes by less than 1\% when the mass splitting of the neutral particles is varied by a factor of 10. Keeping this in mind, we will fix the chargino-to-Higgsino mass splitting to 
$\delta m=355$~MeV, and further adopt $\delta m_N=m_{\chi_2^0}-m_{\chi_1^0} = 20$~MeV for the mass splitting of the two neutral fermions. 
This value is small enough that dimension-5 operators responsible for it do not modify $\delta m$ appreciably but, at the same time, is large enough that---for the range of DM masses considered here---$\chi_2^0\chi_2^0$ cannot be produced for typical Galactic DM velocities $v\sim10^{-3}$.
We refer to \cite{Chun:2012yt} for a comprehensive discussion of the rich Higgsino mass splitting phenomenology. 

The derivation of the factorization of the photon spectrum in 
the $\gamma +X$ final state from DM annihilation near maximal photon 
energy\footnote{Strictly speaking, $\mchi$ refers here to the mass of the DM particle $\chi_1^0$ and not to the mass parameter in \eqref{eq:fullth}. Henceforth, $\mchi$ should be understood as the mass of the DM particle.} $E_\gamma=\mchi$ has been thoroughly discussed in 
Section 2.1 of \cite{Beneke:2019vhz}, and holds for 
generic weakly interacting DM. For the sake of brevity, we will 
therefore provide here the model-specific 
expressions for the Higgsino and recommend the reader to 
consult \cite{Beneke:2019vhz} for theoretical background information. 

Since the Higgsino multiplet has non-vanishing hypercharge, we must 
extend the basis $(\mathcal{O}_{1-3})$ of short-distance annihilation 
operators with three more operators $(\mathcal{O}_{4-6})$. For the 
annihilation process investigated here, only four out of the six 
operators are relevant (see Appendix \ref{app:opBasis} for the 
complete set of operators)\footnote{As in 
\cite{Beneke:2012tg}, the fields $\zeta_v$ and $\eta_v$ are 
destruction operators for non-relativistic particles and 
antiparticles in the $\mathbf{2}$ and $\mathbf{2^*}$ 
representations of SU(2), respectively (before EWSB). 
Note that after interchanging $\zeta_v$ and 
$\eta_v$, the operators remain the same (up to a sign in some 
cases). In contrast to \cite{Beneke:2012tg} we do not include these 
redundant operators, and hence a factor of 2 appears relative to the 
operators defined in \cite{Beneke:2018ssm,Beneke:2019vhz} for the 
wino case. With this convention the tree-level 
matching coefficient of the operator $\mathcal O_2$ is the same 
for Majorana and Dirac DM.} 
\begin{eqnarray}
\mathcal O_1 &=& 2\,\zeta_v^{c\dagger}\Gamma^{\mu\nu}\eta_v\,
\mathcal{A}^B_{\perp c,\mu}(sn_+) \mathcal{A}^B_{\perp \bar{c},\nu}(tn_-)\,,
\label{eq:ophiggsino1}
\\
\mathcal O_2 &=& \zeta_v^{c\dagger}\Gamma^{\mu\nu}
\{T^B,T^C\}\,\eta_v\,
\mathcal{A}^B_{\perp c,\mu}(sn_+) \mathcal{A}^C_{\perp \bar{c},\nu}(tn_-)\,,
\label{eq:ophiggsino2}\\
\mathcal O_4 &=& 2\,\zeta_v^{c\dagger}\Gamma^{\mu\nu}T^C\eta_v\,
[\mathcal{A}^C_{\perp c,\mu}(sn_+) \mathcal{B}_{\perp \bar{c},\nu}(tn_-)+\mathcal{A}^C_{\perp\bar c,\mu}(tn_-) \mathcal{B}_{\perp c,\nu}(sn_+)\ ]\,,
\label{eq:ophiggsino4}\\
\mathcal O_6 &=& 2\,\zeta_v^{c\dagger}\Gamma^{\mu\nu}\,\eta_v\,
\mathcal{B}_{\perp c,\mu}(sn_+) \mathcal{B}_{\perp \bar{c},\nu}(tn_-)\, ,
\label{eq:ophiggsino6}
\end{eqnarray}
where $\mathcal B^\mu$ is the SCET building block for the U(1)$_Y$ gauge field, and the spin-singlet matrix $\Gamma^{\mu\nu}$ is defined in \cite{Beneke:2019vhz}. In accordance with \cite{Beneke:2019vhz} the non-relativistic fields are represented in an unbroken-index notation. However, unlike in the wino case, in the unbroken Higgsino multiplet particles are not their own antiparticles. Thus, we adopt the nomenclature of \cite{Beneke:2012tg} where the $\eta_v$ fields represent particles and $\zeta_v$ the corresponding antiparticles. Additionally,
\begin{align}
\label{eq:opdeg}
\mathcal{O}_2^{\rm Higgsino} = \frac{1}{4} \mathcal{O}_1^{\rm Higgsino},
\end{align}
for Higgsino DM. For $j>1/2$ SU(2) multiplets the 
operators $\mathcal{O}_{1}$ and $\mathcal{O}_{2}$ are linearly 
independent. Hence, the factorization formula can be simplified by 
introducing the Wilson coefficients
\begin{align}
\label{eq:cTilde}
\tilde{C}_1 = \left(C_1 + \frac{1}{4}C_2\right) \bigg|_{j=1/2} \,, \qquad \tilde{C}_4 = C_4|_{j=1/2} \,, \qquad \tilde{C}_6 = C_6|_{j=1/2} \,.
\end{align}

The photon energy spectrum in $\chi_1^0\chi_1^0$ DM annihilation 
can be written as\footnote{The overall factor of 2 
is necessary in the method-2 computation of the Sommerfeld effect 
\cite{Beneke:2014gja} for the annihilation of two identical particles 
to compensate for the method-2 factor $1/(\sqrt{2})^{n_{id}}$ in 
(\ref{eq:logfact}). This factor of 
2 has been missed in \cite{Beneke:2018ssm,Beneke:2019vhz} and hence 
the absolute values of $\langle \sigma v\rangle$ shown in 
Figure~4 in \cite{Beneke:2018ssm} and Figures~3-5 
in \cite{Beneke:2019vhz} (published and arXiv version 1) 
must be multiplied by two.}
\begin{equation}
\label{eq:sommfact}
\frac{d\left(\sigma v_{\rm rel}\right)}{dE_\gamma}= 2 \,
\sum_{I,J}S_{IJ}\Gamma_{IJ}(E_\gamma)\,,\quad I,J=(11)\,,\,(22)\,,
\,(+-)\, , 
\end{equation}
where $S_{IJ}$ captures the Sommerfeld effect, $\Gamma_{IJ}(E_\gamma)$ is the Sudakov-resummed annihilation rate, and the indices $I,J$ run over Higgsino two-particle states $\chi_i\chi_j$.
 The Sommerfeld factor is computed with the leading-order potential in non-relativistic effective theory. The method of computation is discussed in \cite{Beneke:2014gja} and the potential encapsulating the long-range force between the DM particles is given in Appendix \ref{app:pot}. 
For gamma-ray energies $E_\gamma$ in a bin of (intermediate) energy resolution $\Eres \sim \mathcal O(m_W)$ near the endpoint of the spectrum, the Sudakov-resummed annihilation rate can be further factorized:
\begin{eqnarray}
\Gamma_{IJ}(E_\gamma)&=&\frac{1}{(\sqrt{2})^{n_{id}}}\frac{1}{4}
\frac{2}{\pi m_\chi}\sum_{i,j=1,4,6}\tilde{C}_i(\mu)\tilde{C}_j^*(\mu)
Z_\gamma^{WY}(\mu,\nu) \nn\\
&& \times \,\int d\omega{}{}\left(J^{\rm SU(2)}_{\rm int}(4\mchi(\mchi-
E_\gamma-\omega/2),\mu)\,W^{{\rm SU(2)},\,ij}_{IJ,WY}(\omega,\mu,\nu)\right. 
\nonumber\\
&&\hspace{1cm}+\,\left. J^{\rm U(1)}_{\rm int}(4\mchi(\mchi-
E_\gamma-\omega/2),\mu) \,W^{{\rm U(1)},\,ij}_{IJ,WY}(\omega,\mu,\nu)\right).
\label{eq:logfact}
\end{eqnarray}
This factorization formula resembles the one obtained in \cite{Beneke:2019vhz} and the meaning of the functions (and their arguments $\omega$, $\mu$ and $\nu$) is analogous: $\tilde{C}$ are the short-distance coefficients of the hard annihilation processes, $Z_\gamma$ is the photon jet function describing the detected gamma-ray, $J_{\rm int}$ is the jet function describing the unobserved hard-collinear radiation, and $W$ describes the soft radiation. The indices $W,Y=3,4$ appearing in the soft functions and the photon-jet function are adjoint indices of the SU(2)$\otimes$U(1)$_Y$ group (indices $W,Y=4$ denote the U(1)$_Y$ components). Specific expressions of these functions and their resummation are discussed in Appendix \ref{app:nlo}. Relative to the wino case, a new component due to the hypercharge of the Higgsino appears in (\ref{eq:logfact}).

There are three non-relativistic particle-pair indices $I$ and $J$ relevant for $\chi_1^0\chi_1^0\to\gamma+X$ annihilation. Concretely, $I,J=(11),(22),(+-)$ where the indices $(11)$ and $(22)$ refer to the $\chi_1^0\chi_1^0$ and $\chi_2^0\chi_2^0$  two-particle states, respectively, and the remaining index $(+-)$ labels the chargino particle-antiparticle pair. The 
$\chi_1^0\chi_2^0$ state is not relevant, since the  $\chi_1^0\chi_1^0$, 
$\chi_2^0\chi_2^0$ and $\chi^+\chi^-$ states cannot scatter into 
$\chi_1^0\chi_2^0$ prior to the annihilation. 
Since the matrix $\Gamma_{IJ}$ is not dependent on the mass splittings $\delta m$ and $\delta m_N$ at the NLL' accuracy investigated here, the annihilation matrix fulfils the following properties: $\Gamma_{(22)(22)}=\Gamma_{(11)(22)}=\Gamma_{(22)(11)}=\Gamma_{(11)(11)}$ and $\Gamma_{(22)(+-)}=\Gamma_{(11)(+-)}=\Gamma_{(+-)(22)}^*=\Gamma_{(+-)(11)}^*$. The matrix $\Gamma_{IJ}$ has a slightly more complicated structure than for the wino \cite{Beneke:2019vhz} because of the non-vanishing hypercharge of the Higgsino multiplet.

In Appendix \ref{app:factformulas}, we provide a more general 
version of the intermediate resolution factorization theorem, valid 
for general SU(2) multiplets, and more discussion of 
the functions appearing therein. We also provide the corresponding 
Higgsino-DM factorization formula for the narrow 
energy resolution regime $\Eres\sim m_W^2/m_\chi$. 

\section{Mass corrections}
\label{sec:nlp}

The factorization formulas \eqref{eq:sommfact}, \eqref{eq:logfact} 
and \eqref{eq:nrwlogfact}, and the ones obtained in \cite{Beneke:2018ssm,Beneke:2019vhz} are valid up to power corrections, i.e.~terms proportional to $v^2$, $\delta m/\mchi$ or $\lambda\sim m_W/\mchi$ and higher powers. The former two can safely be neglected as $v^2\sim10^{-6}$ for DM annihilation in Milky Way-sized galaxies, and $\delta m/\mchi\sim 10^{-4}$ for the Higgsino and wino models. However, 
$m_W/\mchi\sim0.1\,\times\,$(1~TeV/$\mchi$) is larger than the 
percent-level accuracy of NLL' resummed annihilation rates.
In this section we shall investigate whether such linear 
$\mathcal{O}(\lambda)$, next-to-leading-power mass 
corrections may be important.

In order to assess the impact of power corrections 
on \eqref{eq:sommfact} we calculate the 
$\chi_1^0\chi_1^0\to\gamma\gamma$ amplitude at the lowest, one-loop 
order in the coupling expansion. The computation of $\chi_1^0\chi_1^0\to\gamma Z$ would be similar. We find for the amplitude at 
vanishing relative velocity $v=0$, expanded in $\lambda = m_W/\mchi$ up to $\mathcal{O}(\lambda)$, the expression\footnote{We also repeated this calculation for wino DM and obtained an almost identical result$, i \mathcal{M}_{\rm wino}^{\chi^0 \chi^0 \rightarrow \gamma \gamma}=4i \mathcal{M}_{\rm Higgsino}^{\chi_1^0 \chi_1^0 \rightarrow \gamma \gamma}$, since the only different relevant couplings are 
$\frac{1}{2} \bar{\chi}_1^0\slashed{W}^+\chi^-$ for the Higgsino and 
$\bar{\chi}^0\slashed{W}^+\chi^-$ for the wino, respectively.}
\begin{eqnarray}
i \mathcal{M}^{\chi_1^0 \chi_1^0 \rightarrow \gamma \gamma} &=&
\frac{i e^4}{16\pi^2 s_W^2 m_\chi^2}\,\frac{1}{2} \varepsilon^*_{\mu}(p_3)\varepsilon^*_{\nu}(p_4) \bar{v}(p_2) \left[\gamma^{\mu}, \gamma^{\nu}\right] \slash{p}_3 u(p_1) \nn \\
&&\hspace*{-1.5cm}\times \left[-2\pi\frac{m_\chi}{m_W}\left(1+\frac1{24}\frac{m_W^2}{\mchi^2}\right) -2 +\frac{\pi^2}{4} +(-1 + i\pi)\ln\frac{4m_\chi^2}{m_W^2}+\mathcal{O}\!\left(\frac{m_W^2}{m_\chi^2}\right)\right] .\quad
\label{eq:fixed_order2}
\end{eqnarray}
In order to arrive at this result, we used {\tt feynrules} \cite{Alloul:2013bka} for the model implementation, {\tt FeynArts} \cite{Hahn:2000kx} for the Feynman rules and diagram generation, {\tt FormCalc} \cite{Hahn:1998yk} for amplitude processing, and {\tt PackageX} \cite{Patel:2015tea} for the evaluation of the loop integrals.

The finite and logarithmic pieces of the amplitude are all included in the Sudakov term $\Gamma_{(11)(11)}$ ($\Gamma_{(00)(00)}^{\rm wino}$). This can be verified by removing the $\mchi/m_W$ and $m_W/\mchi$  terms, squaring the resulting amplitude and comparing the result with Eq.~(313) in \cite{Beneke:2019vhz}. The term that is proportional to $\lambda^{-1}=\mchi/m_W$ is the familiar one-loop Sommerfeld-enhancement 
factor. 

\begin{figure}[t]
	\centering
	\hspace*{-0.6cm}
	\includegraphics[width=0.55\textwidth]{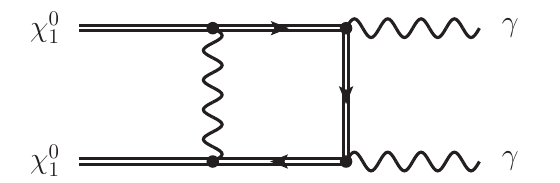}
	\caption{Diagram computed in the potential region to subleading power to subtract the Sommerfeld effect from the fixed-order result.
		\label{fig:sommerfeldDiag}}
\end{figure}

Of particular interest for this discussion is the linear 
mass correction $-2\pi m_\chi/m_W \times m_W^2/(24 m_\chi^2)$. We find that this term is also associated with the non-relativistic dynamics of the problem. We verified this by showing that it arises exclusively from 
expanding the diagram displayed in Figure~\ref{fig:sommerfeldDiag} 
(and the one with the photon lines crossed) to subleading power 
in the potential loop momentum region. 
In the context of non-relativistic 
effective theory, the coefficient $1/24$ originates from 
subleading-power potentials\footnote{These are $\mathcal O(v^2)$ and thus not included in Ref. \cite{Beneke:2019qaa}.} ($-1/8$) and (at the squared-amplitude level) the matrix element of the dimension-8 
derivative S-wave operator $\mathcal P({}^1S_0)$ introduced in \cite{Beneke:2014gja,Hellmann:2013jxa} ($+1/6$).\footnote{For massive mediators, this matrix element is non-vanishing even at zero relative 
momentum. See Eqs.~(119, 120) of \cite{Beneke:2014gja}.} Since in the non-relativistic theory leading-power contributions are $\mathcal O(\lambda^{-1})$, such $\mathcal O(\lambda)$ contributions should be counted as $\mathcal O(\lambda^2)$ corrections to the leading Sommerfeld enhancement. 
Together with the small coefficient $1/24$, which results in a 
$3\times10^{-4}$ correction, we may conclude that we do not 
expect mass corrections to the leading-power 
factorization formula to degrade the percent level accuracy of the
NLL' resummation in the Higgsino mass range of interest.

\section{Numerical results}
\label{sec:results}

As in \cite{Beneke:2018ssm,Beneke:2019vhz} we turn our attention on the cumulative endpoint annihilation rate
\begin{equation}
\label{eq:intspec}
\langle\sigma v\rangle(\Eres)=\int_{\mchi-\Eres}^{\mchi} dE_\gamma\frac{d(\sigma v)}{dE_\gamma}\, .
\end{equation}
For the numerical results given in this section we use the couplings 
at the scale $m_Z = 91.1876 \,\text{GeV}$ in the $\overline{\text{MS}}$ 
scheme as input: $\hat{\alpha}_2(m_Z) = 0.0350009$, $\hat{\alpha}_3(m_Z) = 0.1181$, 
$\hat{s}_W^2(m_Z) = \hat{g}_1^2/(\hat{g}_1^2+\hat{g}_2^2)(m_Z) = 0.222958$, 
$\hat{\lambda}_t(m_Z) = 0.952957$, $\lambda (m_Z) = 0.132944$. The 
$\overline{\text{MS}}$ gauge couplings are in turn computed via one-loop relations 
from $m_Z, m_W = 80.385 \,\text{GeV}$, $\alpha_{\text{OS}}(m_Z) = 1/128.943$. Further, we compute 
the top Yukawa and Higgs self-coupling, which enter our calculation only implicitly 
through the two-loop evolution of the gauge couplings, via tree-level relations from $\overline{m}_t = 163.35 \,\text{GeV}$ (corresponding to the top pole mass 
$173.2 \,\text{GeV}$ at four loops) and $m_H = 125.0 \,\text{GeV}$. The mass splittings are fixed to $\delta m = 355~\text{MeV}$ and $\delta m_N = 20~\text{MeV}$. In \cite{Beneke:2019vhz} two resummation schemes for the intermediate resolution regime were established. The plots and benchmark values in this work were generated with the second resummation scheme and we confirmed that the difference between the two schemes is not larger than $\mathcal{O}(0.1\%)$.

\begin{figure}[t]
	\centering
	\hspace*{-0.6cm}
	\includegraphics[width=0.76\textwidth]{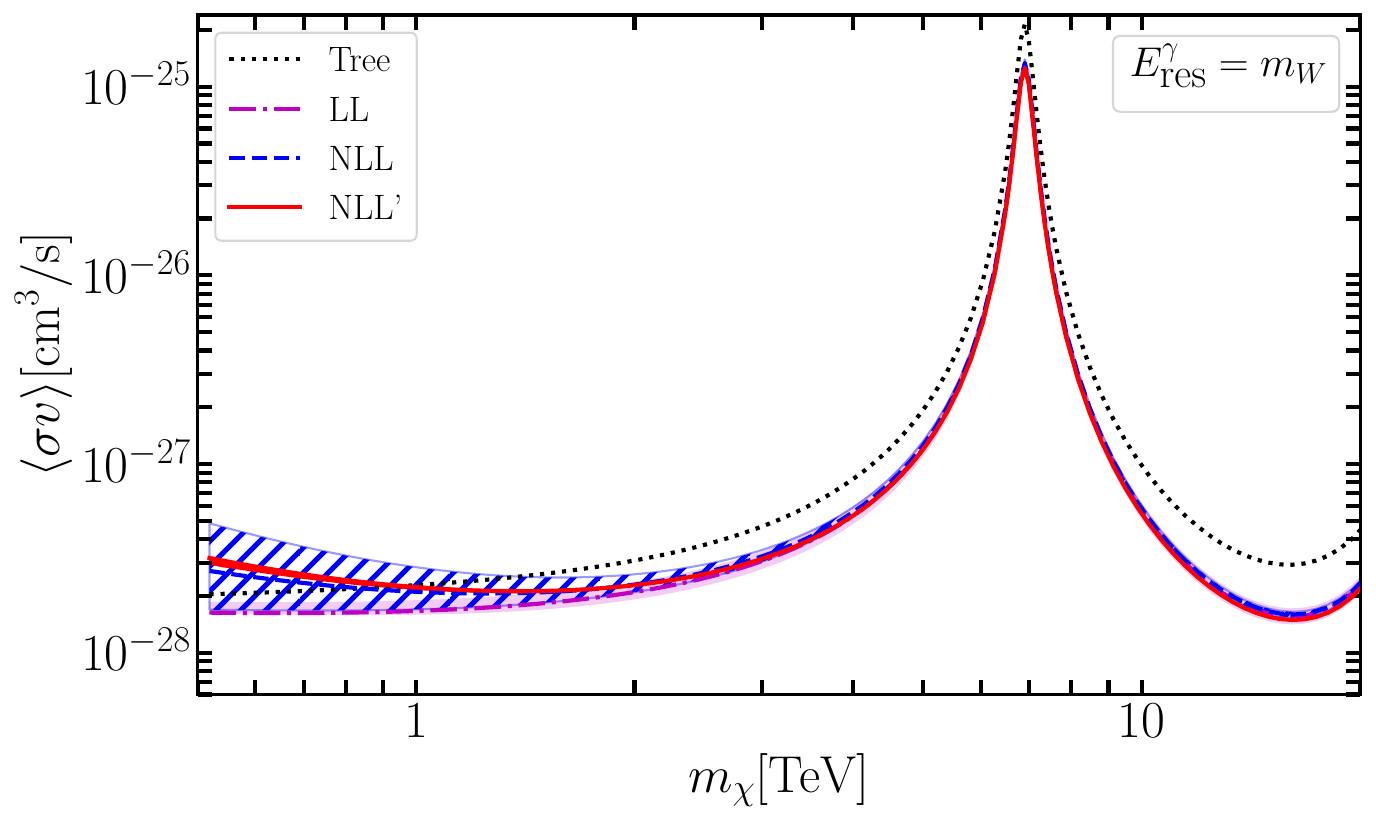} 
	\vskip0.2cm
	\includegraphics[width=0.74\textwidth]{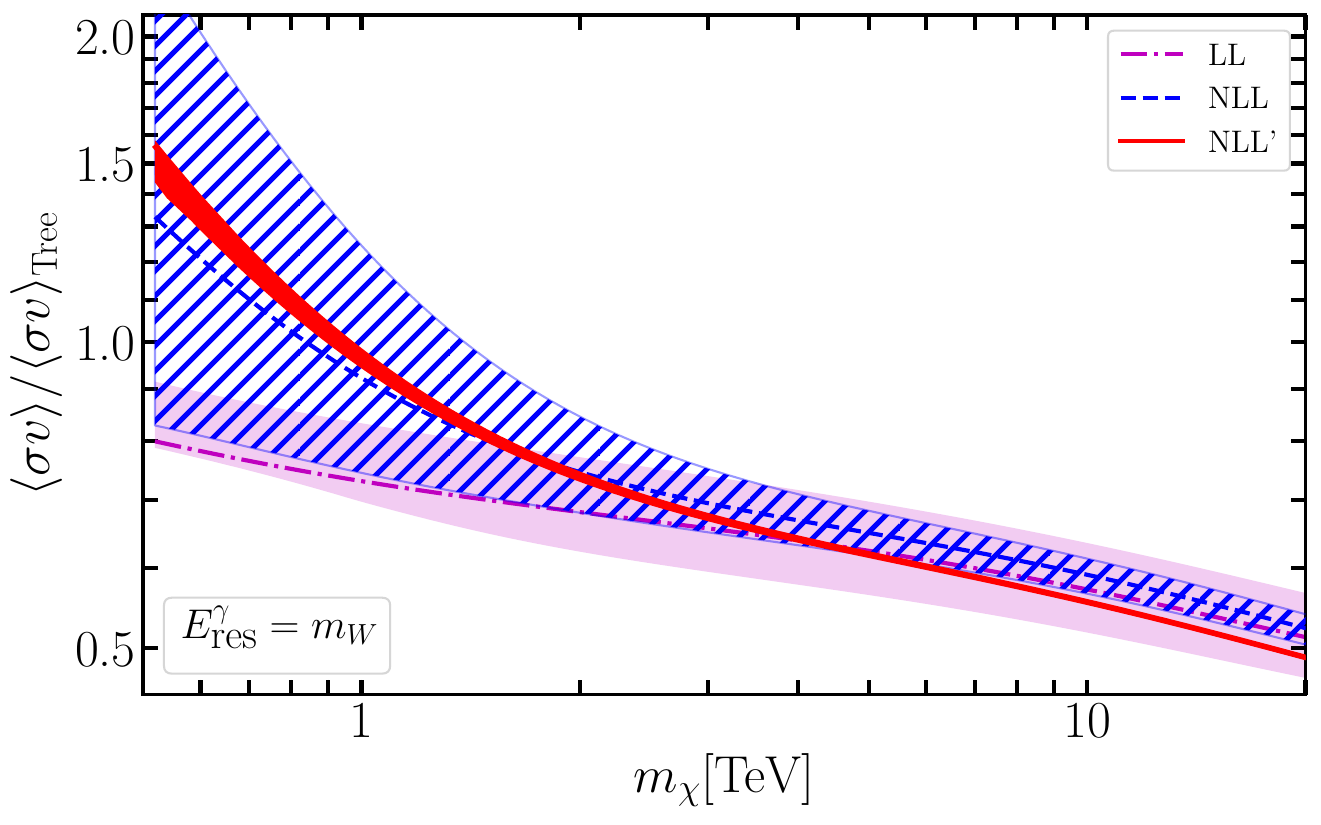}
	\caption{Integrated photon energy spectrum within $E^\gamma_{\text{res}}$ 
		from the endpoint $\mchi$ in the tree (Sommerfeld only) and LL, NLL, NLL' 
		resummed approximations.  The energy resolution is set to $\Eres=m_W$ and the mass splittings are $\delta m = 355~\text{MeV}$ and $\delta m_N = 20~\text{MeV}$. 
		The shaded/hatched bands show the scale variation 
		of the respective approximation. For the NLL' 
		result the theoretical uncertainty is given by the thickness of the red line.
		\label{fig:result}}
\end{figure}

The upper panel in Figure~\ref{fig:result} shows the integrated spectrum $\langle\sigma v\rangle(\Eres)$ plotted as a function of the DM mass $m_\chi$, for the intermediate telescope resolution~\eqref{eq:sommfact} set to $\Eres = m_W$ for definiteness. The displayed DM mass range includes the first Sommerfeld resonance. The different lines refer to the calculation of  $\Gamma_{IJ}$ at tree level (black-dotted), the LL (magenta-dotted-dashed), the NLL (blue-dashed) and the NLL' (red-solid) resummed expression for $\Gamma_{IJ}$. The latter represents the result with the highest accuracy. For better visibility of the resummation effect the lower panel of the Figure~\ref{fig:result} shows the LL, NLL and NLL' resummed annihilation rates normalized to the tree, i.e. Sommerfeld-only result. 
We see that resummation leads to a reduction of the cross section for large mass, as is generally expected for Sudakov resummation. In the low-mass regime around $m_\chi = 1~\tev$ and below, 
however, resummation enhances the NLL and NLL' annihilation rate.

This behaviour can be understood from the following observations. 
1) At large masses the entries of the Sommerfeld matrix $S_{IJ}$ in 
(\ref{eq:sommfact}) have similar magnitude and the sum over $I$, $J$ 
is dominated by the $(+-), (+-)$ term. The annihilation 
rate $\Gamma_{(+-)(+-)}$ in the diagonal 
$\chi^+\chi^-\to\gamma\gamma$ channel starts at tree-level and 
has a standard series of exponentiated negative 
double-logarithmic corrections. On the other hand, for 
masses smaller than about 
1~TeV the Sommerfeld effect is not very effective in mixing 
the various channels, and $S_{(11)(11)} \approx 1$, while 
the other elements are much smaller. Since, however, 
$\Gamma_{(11)(11)}$ is non-vanishing only from 
$\mathcal{O}(\alpha_{1,2}^3)$, all terms in the sum over $I$, $J$ can contribute equally 
to the annihilation rate and partial cancellations may occur. 
2)  Quite generally, at small masses it is not guaranteed that the 
leading logarithms dominate. Moreover, the leading logarithms in 
the neutral annihilation channels are positive, as we discuss 
below. This effect dominates over the negative interference term 
from $\Gamma_{(11)(+-)}$ and the Sudakov suppression of 
$\Gamma_{(+-)(+-)}$, resulting in an enhanced annihilation 
rate at masses below 1~TeV. 

The resummed predictions are shown with theoretical uncertainty bands computed from a parameter scan over the scales, with simultaneous variations of a factor of two of all scales (see also \cite{Beneke:2019vhz}). In the large-mass region the scale dependence decreases as higher orders are successively added from LL to NLL to NLL', but in the low-mass region the error band of the NLL prediction exceeds the error  of the LL result and is very large in absolute terms. For small masses the inclusion of the non-logarithmic one-loop corrections to the hard, jet and soft functions is therefore necessary to gain control over the scale uncertainty, as is done in the NLL' approximation. We then find that the residual theoretical uncertainty at the NLL' order given by the width of the red-solid curve in Figure~\ref{fig:result} is small, and practically negligible for high masses. Numerically, for the two mass values $\mchi = 1~\tev \,(10~\tev)$ the ratio to the Sommerfeld-only rate is $0.730^{+0.102}_{-0.033}$ $(0.571^{+0.063}_{-0.053})$ at LL, $0.922^{+0.323}_{-0.178}$ $(0.590^{+0.022}_{-0.024})$ at NLL and $0.976^{+0.011}_{-0.034}$ $(0.555^{+0.004}_{-0.003})$ at NLL'.  For $\mchi = 1~\tev$, this corresponds to a theoretical uncertainty of $\pm 9\%$ at LL, $\pm 27\%$ at NLL, and only $\pm 2 \%$ at NLL'.

The pattern of scale dependence at large masses is similar to 
the case of the wino  \cite{Beneke:2019vhz}, but the large NLL 
uncertainty at small masses was not seen there. An analysis of the 
analytic expressions for the resummed Higgsino annihilation matrix 
allows us to trace this behaviour to the following feature of 
the Higgsino model. The vanishing of the tree-level amplitude 
for the annihilation of a pair of neutral Higgsinos into 
$\gamma\gamma$ and $\gamma Z$, which implies the vanishing 
of the tree-level annihilation matrix elements in the $(00),(00)$ 
and off-diagonal $(00),(+-)$ ($(+-),(00)$) 
components,\footnote{In the following discussion, (00) always refers 
to the combined neutral states (11), (22).}
is caused by a cancellation between the short-distance coefficients 
of the three operators $\mathcal{O}_{1,4,6}$. This cancellation is 
not preserved by the electroweak renormalization group evolution, 
since the three operators have different anomalous dimensions 
and do not mix.\footnote{Since the photon is not an  electroweak
gauge eigenstate, switching to the mass basis of the operators, 
where the tree-level matching coefficients of the operators 
relevant to the $\gamma\gamma$ and $\gamma Z$ eigenstates is  
zero, does not cancel this effect.} As a consequence there is a double logarithmic 
enhancement proportional to $L^2 = \ln^2(4\mchi^2/m_W^2)$ 
in the one-loop amplitudes despite the absence of a tree amplitude.
Then the lowest non-vanishing order in 
$\Gamma_{(00),(00)}^{\rm NLL}$, which is 
$\mathcal{O}(\alpha_{1,2}^4)$,\footnote{
In general, $\Gamma_{(00),(00)}$ is already non-vanishing at 
$\mathcal{O}(\alpha_{1,2}^3)$ due to the process 
$\chi_i^0\chi_i^0 \to \gamma +W^+ W^-$, which appears 
first in the NLL' approximation.} already carries a $L^4$ enhancement. This is 
in stark contrast to the wino model, where  $\Gamma_{(00),(00)}$ 
contains at most $L^2$ at $\mathcal{O}(\alpha_2^4)$ 
(see Appendix~E of 
\cite{Beneke:2019vhz}). The different logarithmic structure is 
related to the hypercharge of the Higgsino, which allows 
the decay $\chi_i^0\chi_i^0\to ZZ$ at tree level. 

The higher powers of logarithms in the neutral annihilation matrix 
channel, which is  relevant at low masses, is also the origin 
of the large scale dependence of the NLL approximation. 
Defining  $l_{\mu_s} \equiv 
\ln (\mu_s^2/m_W^2)$ and  $l_{\mu_h} \equiv 
\ln (\mu_h^2/4 \mchi^2)$, we have for the first non-vanishing, 
 $\mathcal{O}(\alpha_2^4)$ contribution
\begin{equation}
\label{eq:00twoloop}
\Gamma_{(00),(00)}^{\text{NLL}} = 
\frac{\hat{\alpha}_2^4 \hat{s}_W^2}{64 \pi \mchi^2} 
\left[ \frac{L^4}{4} +L^3 + \# \,L^2 
+ L\left(8\pi^2\,\lmuh -8\pi^2\,\lmus + \ldots \right) \right]
 + \ldots \,,
\end{equation}
where we write explicitly only the terms relevant to the 
discussion. The existence of a $L^4$ term implies that 
the coefficients of $L^2$ and $L$ are dependent on the 
matching scales, consistent with the fact that these 
coefficients are not yet properly summed in the NLL 
approximation. We then find that the large scale dependence 
is caused by the $\pi^2$-enhanced terms shown in 
(\ref{eq:00twoloop}), which in turn stem from the imaginary 
parts of the one-loop anomalous dimensions. The variation of 
these single-logarithmic terms under scale variation is 
larger than the scale-independent part of 
$\Gamma_{(00),(00)}^{\text{NLL}}$, which causes an 
$\mathcal{O}(1)$ scale dependence at small masses.\footnote{This 
serves as a reminder that at $\mchi=0.5$~TeV, the leading 
logarithms in $L$ do not necessarily dominate.} Adding the 
one-loop non-logarithmic terms to the hard, jet and soft 
functions at NLL' removes the scale-dependent terms 
shown in (\ref{eq:00twoloop}), which explains the dramatic 
reduction of the theoretical uncertainty when going from 
NLL to NLL'. The same large scale-dependent terms are 
also present in the LL approximation. 
However, in this case there is an accidental cancellation of 
large scale dependence between the $(00),(00)$ and $(00),(+-)$
contributions to the sum in (\ref{eq:sommfact}), resulting 
in an accidentally small, and highly asymmetric scale 
uncertainty, as seen in Figure~\ref{fig:result}. 
None of these observations 
is relevant to the high-mass regime, where the sum is 
by far dominated by the $(+-),(+-)$ component, which has a 
very small scale dependence already at NLL. Neither are they 
for the wino model, where the large scale dependent 
terms do not appear. Furthermore, the high-mass regime 
sets in earlier for the wino, because the Sommerfeld potential 
is stronger than for the Higgsino.

\begin{figure}[t]
	\centering
	\hspace*{-0.6cm}
	\includegraphics[width=0.66\textwidth]{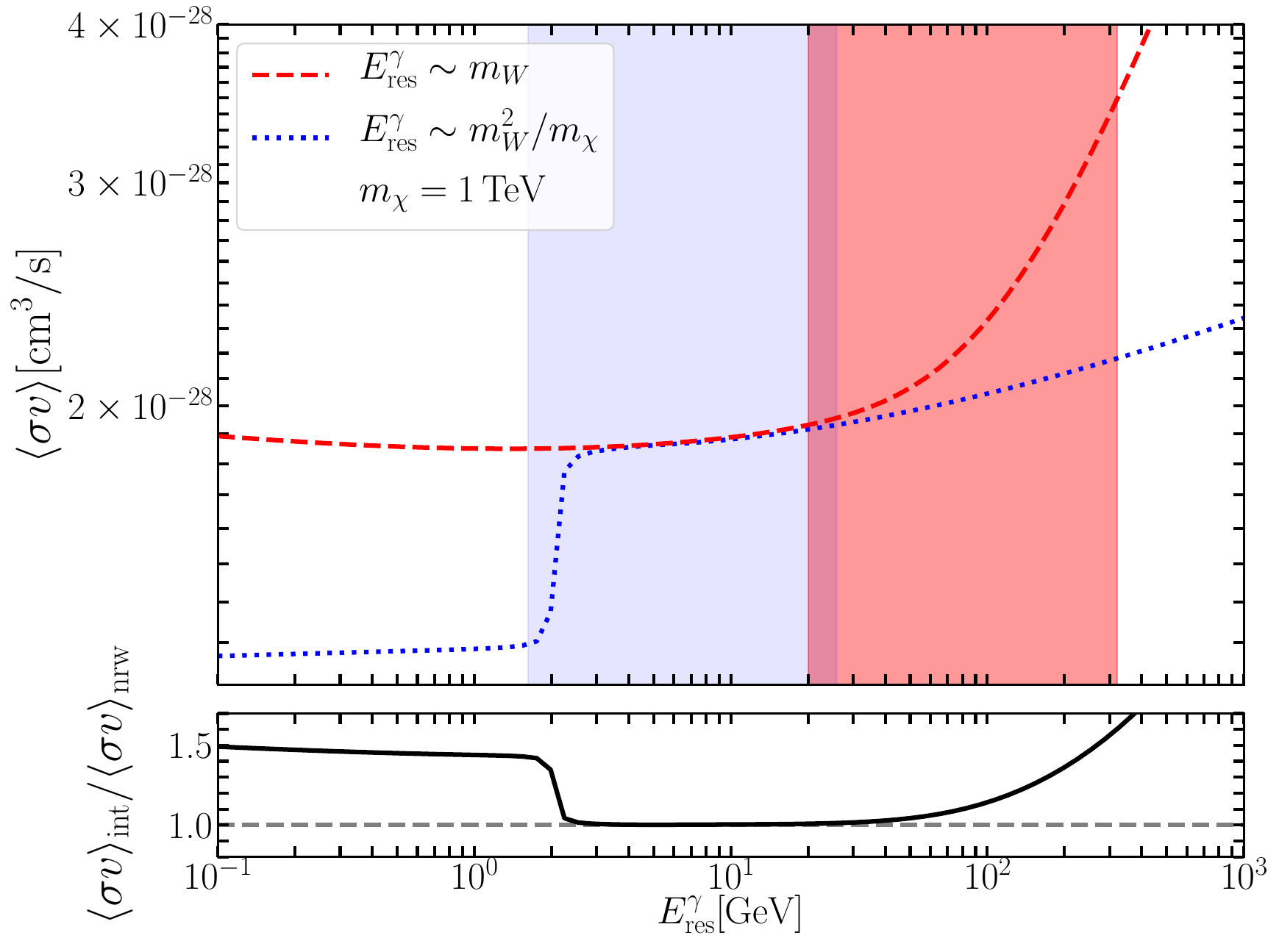} 
	\vskip0.1cm
	\caption{Annihilation cross sections plotted as function of $E_{\text{res}}^\gamma$. The blue-dotted line refers to the narrow resolution. The red-dashed line shows the intermediate resolution cross section. The light-grey (blue) area represents the region of validity of the narrow resolution case and the dark-grey (red) area represents the region of validity of the intermediate resolution case as described in the text. The ratio of the intermediate to narrow resolution annihilation cross section $\langle \sigma v \rangle_{\text{int}}/\langle \sigma v \rangle_{\text{nrw}}$ is added below. The results are shown for a DM mass of $m_\chi = 1 \,\text{TeV}$.
		\label{fig:matching}}
\end{figure}

As for the case of wino DM, it is instructive to investigate whether the narrow $(\Eres \sim m_W^2/m_\chi)$ and the intermediate $(\Eres \sim m_W)$ resolution results can be matched to provide an accurate prediction for the entire range from $\Eres \sim 0$ to $\Eres \approx 4m_W$. In Figure \ref{fig:matching}, we show the annihilation cross sections for the narrow (blue-dotted) and the intermediate (red-dashed) resolution cases, plotted as functions of $\Eres$ for the representative DM mass value $\mchi = 1~\text{TeV}$. We also indicate the natural parametric regions of validity of the narrow resolution (light-grey/blue) and the intermediate resolution (dark-grey/red) computations. The boundaries of these regions are defined by $m_W^2/m_\chi \left[1/4,4\right]$ (narrow resolution) and $m_W \left[1/4,4\right]$ (intermediate resolution).

We observe that there is a wide range of $\Eres$, where the two resolutions match very closely. This implies an accurate theoretical prediction for the photon energy spectrum in DM annihilation in the entire range from $\Eres \sim 0$ to $\Eres \approx 4m_W$. A similar matching was performed in the wino DM case and we found the same degree of matching for the two resolution regimes. An in depth investigation of the structure of the large logarithms was performed for the case of wino DM in \cite{Beneke:2019vhz}, which explained the high degree of matching of the two resolution cases. A similar structure holds for the Higgsino.

There is a steep rise in the narrow resolution cross section, which occurs at $\Eres \simeq m_Z^2/(4\mchi)$. Above this value, the $\gamma Z$ contribution cannot be resolved, resulting in the sharp increase of the semi-inclusive rate. Since the unobserved jet function for the intermediate resolution regime is computed assuming massless particles, it misses this effect. The invariant mass of the narrow resolution unobserved jet function also passes through the $W^+W^-$, $ZH$ and $t\bar{t}$ thresholds, which are however invisible on the scale of the plot.

Due to the high-precision agreement between the two resolution regimes over a wide range of $\Eres$, we can conclude that in Figure \ref{fig:result} the integrated photon energy spectrum of the narrow resolution case would look indistinguishable from the intermediate resolution results, provided the chosen value of $\Eres$ lies somewhat above $m_Z^2/(4\mchi)$ and below $4m_W^2/\mchi$. After performing a similar parameter scan as in the intermediate resolution case to determine the theoretical uncertainty, we find that the reduction of the scale uncertainty with increasing accuracy is comparable for the narrow resolution case.

\section{Conclusions}
\label{sec:concl}
The differential annihilation cross section for Higgsino DM annihilation into $\gamma+X$ near the endpoint was calculated in this paper with the greatest accuracy to date ($\mathcal O$(2\%) at $\mchi=1$~TeV and less at higher masses). This computation systematically includes the resummation of the Sommerfeld effect and electroweak Sudakov logarithms at the NLL' order and is valid for intermediate and narrow energy resolutions. To achieve the above-mentioned accuracy, it is necessary to perform resummation at NLL' accuracy. Our predictions are appropriate for dedicated searches for Higgsino DM (within the thermal production hypothesis) using next-generation gamma-ray telescopes. Electroweak Sudakov resummation decreases the rate of photons by approximately 3\% for a Higgsino mass of 1~TeV. For larger and smaller masses resummation leads to suppression (45\% at 10~TeV) or enhancement, respectively (numbers refer to $\Eres = m_W$).

We also discussed for the first time in the context of indirect DM detection the impact of power-suppressed mass corrections on the endpoint spectrum. In the case of wino and Higgsino DM we find these corrections to the one-loop amplitude to be of $\mathcal O(m_W^2/\mchi^2)$ relative to the leading Sommerfeld effect and thus compatible with the accuracy achieved by NLL' computations. The largest theoretical uncertainty for the spectrum is expected to be associated with NLO corrections to the Sommerfeld potential, which has recently been computed for the wino \cite{Beneke:2019qaa} but is not yet known for Higgsino DM.

The Higgsino annihilation cross sections $\langle\sigma v
\rangle(E^\gamma_{\rm res})\sim 2\times10^{-28}\,{\rm cm}^3/{\rm 
s}$  (for $m_\chi=1\,{\rm TeV}$) predicted here are below the 
limits obtained by existing gamma-ray observatories 
but will be probed by the CTA experiment. Concretely, the
H.E.S.S. experiment quotes an upper limit ($2\sigma$) on $\langle\sigma v\rangle_{\gamma\gamma}$ of $4\times10^{-28}\,{\rm cm}^3/{\rm s}$
 \cite{Abdallah:2018qtu} for their most aggressive assumptions 
on the DM mass distribution of the Milky Way, 
which translates to about  $\langle\sigma v\rangle_{\gamma+X}^{\rm H.E.S.S.\, limit}\approx 8\times10^{-28}\,{\rm cm}^3/{\rm s}$ for the semi-inclusive rate with energy resolution 
$\Eres \approx 100\,$GeV. 
With the help of our results, future measurements of CTA can be 
converted into precise constraints on the Higgsino mass given 
the dark matter profile of the galaxy.

\paragraph{Acknowledgements}
We thank Clara Peset for very useful discussions. This work was supported in part by the DFG Collaborative Research Centre ``Neutrinos and Dark Matter in Astro- and Particle Physics'' (SFB 1258).

\appendix

\section{Sommerfeld potential}
\label{app:pot}
In our numerical evaluations, the method-2 Sommerfeld matrix  $S_{IJ}$ (see \cite{Beneke:2014gja}) is obtained by solving the \schr{} equation with the spin-singlet potential
\begin{equation}
2\delta m_{IJ}+V^{S=0}_{IJ}(r)=\left(
\begin{array}{cccc}
0 & -\frac{\alpha}{4s_W^2 c_W^2}\frac{{\rm e}^{-m_Zr}}{r} & -\frac{\alpha_2}{2\sqrt2}\frac{{\rm e}^{-m_Wr}}{r}\\
-\frac{\alpha}{4s_W^2 c_W^2}\frac{{\rm e}^{-m_Zr}}{r} & 2\delta m_N & -\frac{\alpha_2}{2\sqrt2}\frac{{\rm e}^{-m_Wr}}{r}\\
-\frac{\alpha_2}{2\sqrt2}\frac{{\rm e}^{-m_Wr}}{r} & -\frac{\alpha_2}{2\sqrt2}\frac{{\rm e}^{-m_Wr}}{r} & 2\delta m -\frac{\alpha}{r}-\frac{(1-2c_W^2)^2\alpha}{4s_W^2 c_W^2}\frac{{\rm e}^{-m_Zr}}{r} 
\end{array}
\right).
\end{equation}
The indices are ordered in the following way: $(11)$, $(22)$ and $(+-)$. We added the contribution of the mass-splitting matrix $\delta m_{IJ}$. As mentioned in the main text, there is no interaction between the above three two-particle states and the mixed $(12)$ state.

\section{Factorization formulas}
\label{app:factformulas}

\subsection{Intermediate resolution regime}
\label{app:int}
The derivation of the factorization theorem is independent of whether the DM particle is charged under hypercharge or not. Using the result from \cite{Beneke:2019vhz}, we can thus write the general form of the Sudakov-resummed annihilation rate for the intermediate resolution case as
\begin{align}
\Gamma_{IJ} (E_\gamma,\mu) =& \frac{1}{(\sqrt{2})^{n_{id}}}\frac{1}{4}\frac{2}{\pi m_\chi} \sum_{i,j = 1,2,4,6} C_i(\mu)C_j^*(\mu) Z_\gamma^{YW}(\mu,\nu) \nn \\
&\times\int d\omega J^{XV}_{\rm int}(4\mchi(\mchi -E_\gamma -\omega/2),\mu)W^{ij}_{IJ,VWXY} (\omega) \,.
\label{eq:gammaIJ}
\end{align}
As already discussed in Section \ref{sec:theory}, we now augmented the operator basis to include hard annihilation into the hypercharge gauge boson. The anti-collinear function $Z_\gamma^{YW}$ (photon jet function), the hard-collinear function $J^{XV}_{\rm int}$ (unobserved-jet collinear function) and the soft function $W^{ij}_{IJ,VWXY}$ are generalizations of the corresponding jet and soft functions defined in \cite{Beneke:2019vhz}. In particular, the SU(2)$\otimes$U(1)$_Y$ indices $W$ and $Y$ can now adopt the values 3 and 4, where $3$ refers to the SU(2) gauge boson $W^3$ and 4 to the U(1)$_Y$ gauge boson $B$. It is convenient to split the unobserved jet function into an SU(2) and a U(1)$_Y$ component, as follows:
\begin{align}
J^{XV}_{\rm int} = (\delta^{XV} - \delta^{X4}\delta^{V4}) J^{\rm SU(2)}_{\rm int} + \delta^{X4}\delta^{V4} J^{\rm U(1)}_{\rm int} \,.
\label{eq:jetFuncSplit}
\end{align}
$J^{\rm SU(2)}_{\rm int}$ was already obtained in \cite{Beneke:2019vhz}, while $J^{\rm U(1)}_{\rm int}$ is new to the case of Higgsino DM. The results are presented in Appendix~\ref{app:unobs_jetFunc}. With \eqref{eq:jetFuncSplit} we can reduce the number of relevant indices of the soft function by introducing
\begin{align}
W^{{\rm SU(2)},\, ij}_{IJ,WY}(\omega) &= (\delta^{XV} - \delta^{X4}\delta^{V4})\, W^{ij}_{IJ,VWXY}(\omega) \nonumber \,, \\
W^{{\rm U(1)},\, ij}_{IJ,WY}(\omega) &= \delta^{X4}\delta^{V4}\,W^{ij}_{IJ,VWXY}(\omega) \,.
\end{align}
Here $X,V$ are summed from 1 to 4.
For the pure Higgsino with isospin $j=1/2$, we can also make use of the degeneracy of operators $\mathcal{O}_1$ and $\mathcal{O}_2$ in~\eqref{eq:opdeg}, from which the factorized Sudakov annihilation rate~\eqref{eq:logfact} immediately follows.

\subsection{Narrow resolution regime}
\label{app:nrw}
Assuming the energy resolution $\Eres \sim m_W^2/\mchi$ puts us into the narrow resolution regime. The hierarchy of scales then changes to $\Eres \ll m_W, m_X \sim m_W$, 
which means that the unobserved jet now has collinear rather than hard-collinear virtuality. Furthermore, real soft gauge boson radiation is power suppressed making it convenient to write the soft function as soft Wilson coefficients $D$. A detailed discussion of the factorization theorem for wino DM in the narrow resolution case is provided in \cite{Beneke:2018ssm} and \cite{Beneke:2019vhz}. Since there are no conceptual differences for Higgsino DM, we will not repeat it here. The Sudakov-resummed annihilation rate $\Gamma_{IJ}$ for the narrow resolution case is given by (the Sommerfeld factor $S_{IJ}$ remains the same)
\begin{eqnarray}
\Gamma^{\rm nrw}_{IJ}(E_\gamma)&=&\frac{1}{(\sqrt{2})^{n_{id}}}\frac{1}{4}\frac{2}{\pi m_\chi}\sum_{i,j=1,2,4,6}C_i(\mu)C_j^*(\mu)Z_\gamma^{WY}(\mu,\nu)\nonumber\\ 
&&\times \,D^i_{I,VW}(\mu,\nu)D^{j*}_{J,XY}(\mu,\nu) J_{\rm nrw}^{VX}(4\mchi(\mchi 
-E_\gamma),\mu,\nu)\ ,
\label{eq:nrwlogfact}
\end{eqnarray}
where the SU(2)$\otimes$U(1)$_Y$ indices summed over 
are $V,W,X,Y=3,4$.

\section{NLO expressions}
\label{app:nlo}
In this appendix we provide all expressions that are required for the evaluation of the factorization formulas \eqref{eq:logfact} and \eqref{eq:nrwlogfact} at the one-loop order as is needed for the NLL' resummation. We also provide the necessary results for resumming the various functions. Since there are no conceptual differences with respect to the resummation in the wino DM case, we refer to \cite{Beneke:2019vhz} for an in depth discussion of the systematics of resummation and will keep the present discussion rather brief. Also, all function definitions have already been presented in \cite{Beneke:2018ssm} and \cite{Beneke:2019vhz} and will not be repeated here.

\subsection{Operator basis and Wilson coefficients}
\label{app:opBasis}

In the general case where the field $\chi$ in \eqref{eq:fullth} is a $(2j+1)$-plet of SU(2) with non-vanishing hypercharge, the operators allowed by the symmetries of the SM are
\begin{eqnarray}
\mathcal O_1 &=& 2\,\zeta_v^{c\dagger} \Gamma^{\mu\nu}\eta_v\,
\mathcal{A}^B_{\perp c,\mu}(sn_+) \mathcal{A}^B_{\perp \bar{c},\nu}(tn_-)\,,
\label{eq:opbasis1}
\\
\mathcal O_2 &=&  \zeta_v^{c\dagger}\Gamma^{\mu\nu}
\{T^B,T^C\}\eta_v\,
\mathcal{A}^B_{\perp c,\mu}(sn_+) \mathcal{A}^C_{\perp \bar{c},\nu}(tn_-)\,,
\label{eq:opbasis2}\\
\mathcal O_3 &=& 2\,\zeta_v^{c\dagger}\sigma^\alpha(n_{-\,\alpha}-n_{+\,\alpha})
T^A\eta_v\,
\epsilon^{ABC}\mathcal{A}^{\mu\,B}_{\perp c}(sn_+) \mathcal{A}^C_{\perp \bar{c},\mu}(tn_-)\,,
\label{eq:opbasis3}\\
\mathcal O_4 &=& 2\,\zeta_v^{c\dagger}\Gamma^{\mu\nu}T^C\eta_v\,
[\mathcal{A}^C_{\perp c,\mu}(sn_+) \mathcal{B}_{\perp \bar{c},\nu}(tn_-)+\mathcal{A}^C_{\perp\bar c,\mu}(tn_-) \mathcal{B}_{\perp c,\nu}(sn_+)]\,,
\label{eq:opbasis4}\\
\mathcal O_5 &=& 2\,\zeta_v^{c\dagger}\sigma^\alpha(n_{-\,\alpha}-n_{+\,\alpha})
T^C\eta_v\,[\mathcal{A}^{\mu\,C}_{\perp c}(sn_+) \mathcal{B}_{\perp \bar{c},\mu}(tn_-)-\mathcal{A}^C_{\perp\bar c,\mu}(tn_-)\mathcal{B}_{\perp c,\nu}(sn_+)]\,,\qquad
\label{eq:opbasis5}\\
\mathcal O_6 &=& 2\,\zeta_v^{c\dagger}\Gamma^{\mu\nu}\eta_v\,
\mathcal{B}_{\perp c,\mu}(sn_+) \mathcal{B}_{\perp \bar{c},\nu}(tn_-)\, .
\label{eq:opbasis6}
\end{eqnarray}
The derivation of this operator basis is as in \cite{Beneke:2019vhz}. As argued there, operator $\mathcal O_3$ is irrelevant for the $\chi_1^0\chi_1^0\to\gamma+X$ process. We also find that $\mathcal O_5$ is irrelevant as it is a spin-triplet operator similar to $\mathcal O_3$. The spin-triplet operators do not contribute, since there is no spin-triplet   $\chi_1^0 \chi_1^0$ initial state, and the Sommerfeld-enhanced scattering prior to annihilation does not change the spin. 

The one-loop short-distance coefficients of operators $\mathcal O_{1}$ and $\mathcal O_{2}$ have already been obtained in \cite{Beneke:2019vhz}. For completeness we provide the Wilson coefficients for the entire operator basis~\eqref{eq:opbasis1}-\eqref{eq:opbasis6}. The Wilson coefficients are computed from matching the full-theory amplitude to the effective theory amplitude. The Feynman diagrams are calculated in the unbroken SU(2)$\otimes$U(1)$_Y$ gauge theory. For general $j$ and $Y$, the coefficients are given by
\begin{align}
C_1(\mu) =&\, \, \frac{\hat{g}^4_2(\mu)}{16 \pi^2}\, c_2(j)\, 
\Big[ (2-2 i \pi) \ln\frac{\mu^2}{4 m^2_\chi}
- \Big(4 - \frac{\pi^2}{2}\Big)\Big]\,, \nn\\
C_2(\mu) =&\, \hat{g}_2^2(\mu) + \frac{\hat{g}_2^2(\mu) \,\hat{g}_1^2(\mu) Y^2}{16\pi^2}\left(\frac{\pi^2}{2}-10\right) \nn\\
&+ \frac{\hat{g}^4_2(\mu)}{16 \pi^2}\Big[16 - \frac{\pi^2}{6} + c_2(j) \Big(\frac{\pi^2}{2}-10\Big) - (6-2i\pi) \ln\frac{\mu^2}{4 m^2_\chi}-2 \ln^2\frac{\mu^2}{4 m^2_\chi}\Big] \,, \nn \\
C_3(\mu)|_{\rm Dirac} =&\, \frac{\hat{g}_{2}^2(\mu) \,\hat{g}_{1}^2(\mu) Y^2}{16\pi^2}  \left(- 4 + 2\pi^2 - 16\ln 2\right) + \frac{\hat{g}_{2}^4(\mu)}{16\pi^2} \left[ \frac{20}{3} -2\pi^2 +8\ln 2 \right. \nn \\ 
&\left.+ \,c_2(j)\left(-4 + 2\pi^2 -16\ln 2 + (2 j+1)\left(\frac{26}{9}-\frac{\pi^2}{3}+\frac{2}{9}n_G\right)\right)\right] \nn \,, \\
C_4(\mu) =&\, -\hat{g}_2(\mu) \hat{g}_1(\mu) Y - \frac{\hat{g}_2(\mu) \,\hat{g}_1^3(\mu) Y^3}{16\pi^2} \left( \frac{\pi^2}{2} - 10 \right) \nn \\
&\,- \frac{\hat{g}_2^3(\mu) \hat{g}_1(\mu) Y}{16\pi^2} \left[\frac{\pi^2}{6} + 6 + c_2(j)\left( \frac{\pi^2}{2} - 10 \right) -2\ln\frac{\mu^2}{4\mchi^2} -\ln^2\frac{\mu^2}{4\mchi^2}\right] \,, \nn \\
C_5(\mu) =&\, 0 \,, \nn \\
C_6(\mu) =&\, \hat{g}_1^2(\mu) Y^2 + \frac{\hat{g}_2^2(\mu) \hat{g}_1^2(\mu) Y^2}{16\pi^2} \, c_2(j)\left( \frac{\pi^2}{2} -10 \right) + \frac{\hat{g}_1^4(\mu) Y^4}{16 \pi^2} \left( \frac{\pi^2}{2} -10 \right) \,.
\label{eq:WilsonCoeffsFin}
\end{align}
Here $c_2(j) = j(j+1)$ is the SU(2) Casimir of the isospin-$j$ representation and $n_G = 3$ is the number of fermion generations.\footnote{The matching coefficient for $\mathcal{O}_3$ was previously given for $j=1$ and $Y=0$, i.e.~the pure wino, in \cite{Ovanesyan:2016vkk}. The result given there differs from ours. We attribute this difference to an opposite sign in one of the diagrams in \cite{Ovanesyan:2016vkk} -- namely $T_{5b}$, and missing mass renormalization (counterterm) diagrams. However, as mentioned before, $\mathcal{O}_3$ does not contribute to the annihilation rate into photons.} Note that $C_3(\mu)|_{\rm Dirac}$ in~\eqref{eq:WilsonCoeffsFin} is specific to a Dirac SU(2)$\otimes$ U(1)$_Y$ multiplet. 
If, on the other hand, we assume a Majorana multiplet, we find 
\begin{align}
C_3 (\mu)|_{\rm Majorana} =&\, \frac{\hat{g}_2^4(\mu)}{16\pi^2} \left[ \frac{20}{3} -2\pi^2 +8\ln 2 \right. \nn \\
&\left.+ c_2(j)\left(-4 + 2\pi^2 -16\ln 2 + (2j+1)\left(\frac{4}{3}-\frac{\pi^2}{6}+\frac{2}{9}n_G\right)\right)\right] \,.
\end{align}

Let us comment on a subtlety in the computation of $C_3$. 
Naively, one would expect that there should be no counterterm contributions as the tree-level contribution cancels. 
However, as also observed for the corresponding quarkonium calculation in QCD \cite{Beenakker:2015mra}, the vanishing tree-level result stems from a cancellation between an $s$-channel diagram and the $t/u$-channel diagrams. Since DM mass counterterm insertions exist only for the $t$- and $u$-channel diagram, a mass renormalization contribution survives, which is required to obtain a finite result for $C_3$.
One might wonder how this result can be obtained in bare perturbation theory, as the tree-level result vanishes, and hence there is apparently no bare mass to substitute by the renormalized one. In bare perturbation theory, the appearance of the counterterm is related to a subtlety in the matching procedure. When performing one-loop on-shell matching we must set the mass of the external particles to their renormalized on-shell values at the corresponding loop order. The tree-level diagrams then contain 
the bare mass from the explicit mass in the propagator and the 
renormalized mass from momenta after applying on-shell kinematics. 
The above mentioned cancellation then leaves over the one-loop 
difference between bare and renormalized mass in the $t$- and 
$u$-channel tree diagrams and the same result as before is recovered in bare perturbation theory. 

The operator $\mathcal{O}_5$ is irrelevant for $\chi_1^0\chi_1^0$ annihilation, but we also find its coefficient to be zero at the one-loop order. This is a consequence of the Landau-Yang theorem. Although it does not hold in non-abelian gauge theories (see, for instance, \cite{Beenakker:2015mra, Cacciari:2015ela}), the violation arises due to the fact that the final state bosons carry an internal quantum number and structures can be constructed involving the group structure constant. At least to the one-loop order, there are no such Feynman diagram structures for $\mathcal{O}_5$ as one of the final-state gauge bosons is abelian.

Next we discuss the renormalization-group (RG) evolution of the short-distance coefficients. Recall that we are considering the annihilation process $\chi_1^0 \chi_1^0 \rightarrow \gamma + X$ for Higgsino DM, for which the operators $\mathcal{O}_3$ and $\mathcal{O}_5$ are irrelevant. Hence, we will disregard $C_3(\mu)$ and $C_5(\mu)$ from now on. The evolution of the vector $\tilde{C} = (\tilde{C}_1,\tilde{C}_4,\tilde{C}_6)$ defined in~\eqref{eq:opdeg} and~\eqref{eq:cTilde} is given by
\begin{align}
\tilde{C}(\mu) =
\begin{pmatrix}
U^{(0)}_1(\mu_h,\mu) & 0 & 0 \\
0 & U^{(1)}_4(\mu_h,\mu) & 0 \\
0 & 0 & U^{(0)}_6(\mu_h,\mu)
\end{pmatrix}
\tilde{C}(\mu_h) \,.
\label{eq:resummedWilson}
\end{align}
The evolution factors satisfy the RG equation
\begin{align}
\frac{d}{d\ln\mu} U^{(J)}_i(\mu_h,\mu) = \left(\Gamma^{(J)}_{\text{SU(2)},i} + \Gamma_{\text{U(1)},i}\right) U^{(J)}_i(\mu_h,\mu) \,,
\label{eq:wilsonRGE}
\end{align}
with $\Gamma^{(J)}_{\text{SU(2)},i}$ and $\Gamma_{\text{U(1)},i}$ of the form \cite{Beneke:2009rj}
\begin{eqnarray}
\Gamma_{\text{SU(2)},i}^{(J)} &=& \frac{1}{2}\gamma_{\rm cusp} \left[c_2(\text{ad})\, n_{i,\text{SU(2)}} \left(\ln\frac{4m_\chi^2}{\mu^2} -i\pi \right) + i\pi c_2(J)\right] + \gamma_{\rm ad}\, n_{i,\text{SU(2)}} + \gamma_{H,s}^J\,, \nn \\
\Gamma_{\text{U(1)},i} &=& \gamma_{\rm U(1)}\, n_{i,\text{U(1)}} \,,
\label{eq:anDimsWilson}
\end{eqnarray}
where $n_{i,\text{SU(2)}}$ and $n_{i,\text{U(1)}}$ give the number of SU(2) and U(1) gauge fields in operator $\mathcal{O}_i$. Eq.~\eqref{eq:wilsonRGE} is solved numerically to obtain the resummed short-distance coefficients~\eqref{eq:resummedWilson}. Most anomalous dimensions in~\eqref{eq:anDimsWilson} have already been given in \cite{Beneke:2018ssm} and will not be repeated here. The new ones relevant for the evolution of $\tilde{C}_4(\mu)$ and $\tilde{C}_6(\mu)$ are
\begin{eqnarray}
\gamma_{\rm U(1)} (\hat{\alpha}_1) &=& \gamma_{\rm U(1)}^{(0)} \frac{\hat{\alpha}_1(\mu)}{4\pi} + \mathcal{O}(\hat{\alpha}_1^2(\mu)) \,, \\
\gamma_{\rm U(1)}^{(0)} &=& - \beta_{0, \text{U(1)}} = \frac{1}{6} + \frac{20}{9}n_G \,.
\end{eqnarray}

\subsection{Photon jet function}
\label{app:photon_jetFunc}

\begin{figure}[t]
	\centering
	\hspace*{-0.6cm}
	\includegraphics[width=0.80\textwidth]{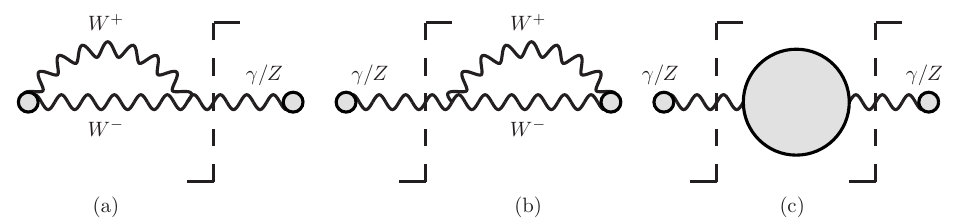}
	\caption{Diagrams contributing to the photon jet functions.
		\label{fig:photon_jet}}
\end{figure}

In Figure~\ref{fig:photon_jet} we show all diagrams relevant for the computation of the photon jet functions. Diagrams (a) and (b) give identical contributions and originate from Wilson lines. The self-energy diagram (c) includes the contributions from all SM particles in the loop. The result for the photon jet function $Z_\gamma^{33}$ was already given in \cite{Beneke:2018ssm,Beneke:2019vhz} and will not be repeated here. For the case of Higgsino DM, which has non-vanishing hypercharge, we also need to take into account the index combinations $Z_\gamma^{34}$, $Z_\gamma^{43}$ and $Z_\gamma^{44}$, which can be obtained by adapting $Z_\gamma^{33}$ accordingly. To do so, it is helpful to remark that only diagram (a) ((b)) contributes to $Z_\gamma^{34}$ ($Z_\gamma^{43}$), while $Z_\gamma^{44}$ does not receive contributions from (a) and (b). Hence, the Wilson line part of $Z_\gamma^{34}$ and $Z_\gamma^{43}$ is multiplied with a factor of $1/2$ with respect to $Z_\gamma^{33}$ while $Z_\gamma^{44}$ does not have a Wilson line contribution at all. To make the computation more transparent, it is convenient to write down the rotation from the weak basis to the mass basis
\begin{align}
\begin{pmatrix}
W^1 \\
W^2 \\
W^3 \\
B
\end{pmatrix}
=
\begin{pmatrix}
\frac{1}{\sqrt{2}} & \frac{1}{\sqrt{2}} & 0 & 0 \\
\frac{i}{\sqrt{2}} & \frac{-i}{\sqrt{2}} & 0 & 0 \\
0 & 0 & \hat{s}_W(\mu) & \hat{c}_W(\mu) \\
0 & 0 & -\hat{c}_W(\mu) & \hat{s}_W(\mu) \\
\end{pmatrix}
\begin{pmatrix}
W^+ \\
W^- \\
\gamma \\
Z
\end{pmatrix}
\label{eq:rotMatrix}
\end{align}
of the gauge fields. One can now use~\eqref{eq:rotMatrix} to compute the remaining photon jet functions $Z_\gamma^{34}$, $Z_\gamma^{43}$ and $Z_\gamma^{44}$ by carefully analyzing the position of the cut in diagrams (a), (b) and (c) and by keeping track of whether the $\gamma$ or the $Z$ boson originated from a $W^3$ or a $B$ boson in the unbroken theory. We find
\begin{align}
Z^{34}_\gamma\left(\mu,\nu\right) &= -\hat{s}_W(\mu) \hat{c}_W(\mu)\bigg[1 -\frac{\hat{g}_2^2(\mu)}{(4\pi)^2}\bigg\{- {8}\ln\frac{m_W}{\mu} \ln\frac{2m_\chi}{\nu} + 4\ln\frac{m_W}{\mu} \bigg\} \, \nn\\
{}& +\frac{1}{2}\left(\frac{\hat{g}_1^2(\mu) \hat{c}_W^2(\mu)}{(4\pi)^2} + \frac{\hat{g}_2^2(\mu) \hat{s}_W^2(\mu)}{(4\pi)^2}\right) \nn \\
&\hspace{2cm}\times \bigg\{\frac{80}{9} \bigg(2\ln\frac{m_Z}{\mu}-\frac{5}{3}\bigg) + \frac{32}{9} \ln\frac{m_t}{\mu}  +\frac{2}{3}-6\ln\frac{m_W}{\mu} \bigg\} \nn \\
&- \left(\frac{\hat{g}_2^2(\mu)}{(4\pi)^2} - \frac{\hat{g}_1^2(\mu) }{(4\pi)^2}\right) \frac{m_W^2}{m_Z^2}\bigg(-4\ln\frac{m_W}{\mu}\bigg) -\Delta\alpha \bigg] \,,
\end{align}
\begin{align}
Z^{43}_\gamma\left(\mu,\nu\right) = Z^{34}_\gamma\left(\mu,\nu\right) \,,
\end{align}
\begin{align}
\label{eq:photonJet44}
Z^{44}_\gamma\left(\mu\right) &= \hat{c}_W^2(\mu) \bigg[1 +\frac{\hat{g}_1^2(\mu) \hat{c}_W^2(\mu)}{(4\pi)^2}\bigg\{   \frac{80}{9} \bigg(2\ln\frac{m_Z}{\mu}-\frac{5}{3}\bigg)+ \frac{32}{9} \ln\frac{m_t}{\mu} \nn\\
{}&\hspace{2cm} +\frac{2}{3} -6 \ln\frac{m_W}{\mu}\bigg\} - \frac{\hat{g}_1^2(\mu)}{(4\pi)^2}\ 8\frac{m_W^2}{m_Z^2} \ln\frac{m_W}{\mu}  -\Delta\alpha \bigg] \,.
\end{align}
The discussion of the RG and rapidity RG evolution will be limited to $Z_\gamma^{34}=Z_\gamma^{43}$ and $Z_\gamma^{44}$. The resummation of $Z_\gamma^{33}$ was discussed in \cite{Beneke:2019vhz} in great detail, which allows us to keep the following analysis rather brief. The RG equations are given by
\begin{eqnarray}
\label{eq:rgePhotonJets}
\frac{d}{d\ln\mu} Z_\gamma^{34} (\mu,\nu) &=& \gamma_{Z_\gamma^{34}}^\mu Z_\gamma^{34}(\mu,\nu) \,, \nn \\
\frac{d}{d \ln\mu} Z_\gamma^{44} (\mu) &=& \gamma_{Z_\gamma^{44}} Z_\gamma^{44} (\mu) \,,
\end{eqnarray}
with the anomalous dimensions
\begin{eqnarray}
\gamma_{Z_\gamma^{34}}^\mu &=& 2 \gamma_{\rm cusp} \ln\frac{\nu}{2m_\chi} + \gamma_{Z_\gamma}^{\rm SU(2)} + \gamma_{Z_\gamma}^{\rm U(1)} \,, \nn \\
\gamma_{Z_\gamma^{44}} &=& 2 \gamma_{Z_\gamma}^{\rm U(1)} \,.
\end{eqnarray}
The anomalous dimensions $\gamma_{Z_\gamma}^{\rm SU(2)}$ 
and  $\gamma_{Z_\gamma}^{\rm U(1)}$ are given by
\begin{equation}
\gamma_{Z_\gamma}^{\text{SU(2)}} = 
\beta_{0,\text{SU(2)}} \, \frac{\hat{\alpha}_2(\mu)}{4\pi} 
+\ldots, \qquad 
\gamma_{Z_\gamma}^{\text{U(1)}} = \beta_{0,\text{U(1)}}  
\,\frac{\hat{\alpha}_1(\mu)}{4\pi} 
+\ldots\,.
\end{equation}
Since $Z_\gamma^{44}$ is independent of the rapidity scale $\nu$, we only need to consider the rapidity RG equation for $Z_\gamma^{34}$. It reads
\begin{align}
\frac{d}{d\ln\nu} Z_\gamma^{34} (\mu,\nu) = \gamma_{Z_\gamma^{34}}^\nu Z_\gamma^{34}(\mu,\nu) \,,
\end{align}
where the rapidity anomalous dimension $\gamma_{Z_\gamma^{34}}^\nu$ is given by
\begin{align}
\label{eq:rrgPhotonJets}
\gamma_{Z_\gamma^{34}}^\nu = \frac{\hat{\alpha}_2(\mu)}{4\pi}2\gamma_{\rm cusp}^{(0)} \ln\frac{\mu}{m_W} \,.
\end{align}
Solving~\eqref{eq:rgePhotonJets} and~\eqref{eq:rrgPhotonJets}, we can compute the resummed photon jet functions
\begin{align}
Z_\gamma^{34}(\mu_f,\nu_f) =& \exp\left[\,\int_{\ln\mu_i}^{\ln\mu_f} d\ln\mu \,\left(2 \gamma_{\rm cusp} \ln\frac{\nu_f}{2m_\chi} + \gamma_{Z_\gamma}^{\rm SU(2)} + \gamma_{Z_\gamma}^{\rm U(1)}\right)\right] \nn \\
&\times \exp\left[\frac{\gamma_{\rm cusp}^{(0)}}{2\beta_{0,\text{SU(2)}}} \ln\frac{\hat{\alpha}_2(\mu_i)}{\hat{\alpha}_2(m_W)} \ln\frac{\nu_i^2}{\nu_f^2}\right] Z_\gamma^{34}(\mu_i,\nu_i)\,, \nn \\
Z_\gamma^{44}(\mu_f) =& \exp\left[\,\int_{\ln\mu_i}^{\ln\mu_f} d\ln\mu \,\gamma_{Z_\gamma^{44}}\right] Z_\gamma^{44}(\mu_i) \,,
\label{eq:resummedPhotonJets}
\end{align}
where the integrals in the exponents are computed numerically.

\subsection{Unobserved jet function}
\label{app:unobs_jetFunc}

\subsubsection{Intermediate resolution}

In addition to the SU(2) gauge-boson jet function $J^{\rm SU(2)}_{\rm int}(p^2,\mu)$ obtained at NLO in \cite{Beneke:2019vhz}, the factorization formula \eqref{eq:logfact} also requires its U(1)$_Y$ counterpart. It is given by
\begin{align}
J^{\rm U(1)}_{\rm int}(p^2) = \delta(p^2) + \frac{\hat{g}_1^2(\mu)}{16\pi^2} \bigg\{ &\delta(p^2) \left( - \frac{104}{9}\right) + \left[\frac{1}{p^2}\right]_*^{\left[\mu^2\right]} \left(\frac{1}{6}+\frac{20}{9}n_G\right) \bigg\}\,.
\label{eq:fixedOrderJU1}
\end{align}
The Laplace transform of $J_{\rm int}^{\rm U(1)}(p^2)$ is defined as
\begin{align}
j^{\rm U(1)}_{\rm int} \left(\ln\frac{\tau^2}{\mu^2},\mu\right) = \int_{0}^{\infty} dp^2 e^{-l p^2} J^{\rm U(1)}_{\rm int} (p^2, \mu) \,,
\end{align}
where $l = 1/(e^{\gamma_E} \tau^2)$ and the explicit result reads
\begin{align}
\label{eq:laplaceTransfjU1}
j^{\rm U(1)}_{\rm int} \left(\ln\frac{\tau^2}{\mu^2},\mu\right) = 1 + \frac{\hat{\alpha}_1(\mu)}{4\pi} \left[\left(\frac{1}{6} + \frac{20}{9}n_G\right) \ln\frac{\tau^2}{\mu^2} - \frac{104}{9}\right] \,.
\end{align}
The corresponding RG equation is the ordinary differential equation
\begin{align}
\frac{d}{d\ln\mu} j^{\rm U(1)}_{\rm int} \left(\ln\frac{\tau^2}{\mu^2},\mu\right) = \gamma_{J^{\rm U(1)}}^\mu \,j^{\rm U(1)}_{\rm int} \left(\ln\frac{\tau^2}{\mu^2},\mu\right) \,,
\label{eq:recoilJetU1rge}
\end{align}
with the Laplace-space anomalous dimension
\begin{align}
\gamma_{J^{\rm U(1)}}^\mu = -2 \gamma_{J^{\rm U(1)}} \,.
\end{align}
$\gamma_{J^{\rm U(1)}}$ is needed at the one-loop order for NLL$'$ resummation:
\begin{align}
\gamma_{J^{\rm U(1)}} = \frac{\hat{\alpha}_1(\mu)}{4\pi} \gamma_{J^{\rm U(1)}}^{(0)} + \mathcal{O}(\hat{\alpha}_1^2) \,, \quad \gamma_{J^{\rm U(1)}}^{(0)} = -\beta_{0, \text{U(1)}} \,.
\end{align}
The RG equation~\eqref{eq:recoilJetU1rge} is solved by
\begin{align}
j^{\rm U(1)}_{\rm int} \left(\ln\frac{\tau^2}{\mu^2},\mu\right) =&\, U(\mu_j,\mu) \,j^{\rm U(1)}_{\rm int} \left(\ln\frac{\tau^2}{\mu^2},\mu\right) \nn \\
=&\, \exp\left[-2\int_{\ln\mu_j}^{\ln\mu} d\ln\mu' \,\gamma_{J^{\rm U(1)}}(\hat{\alpha}_1(\mu')) \right] j^{\rm U(1)}_{\rm int} \left(\ln\frac{\tau^2}{\mu^2},\mu_j\right) \,,
\label{eq:evolUjet}
\end{align}
where $\mu_j \sim \sqrt{2m_\chi m_W}$ is the natural scale of the hard-collinear jet function. To return to momentum space, we perform a standard inverse Laplace transform (remembering that $\tau^2=1/(e^{\gamma_E}l)$). We find for the resummed unobserved jet function $J^{\rm U(1)}_{\rm int}$ 
\begin{align}
J^{\rm U(1)}_{\rm int} (p^2, \mu) = U(\mu_j,\mu) J^{\rm U(1)}_{\rm int} (p^2, \mu_j) \,,
\label{eq:resummedJU1}
\end{align}
where $U(\mu_j,\mu)$ is taken from~\eqref{eq:evolUjet} and $J^{\rm U(1)}_{\rm int} (p^2, \mu_j)$ on the right-hand side of~\eqref{eq:resummedJU1} is given by~\eqref{eq:fixedOrderJU1}. The integral in the evolution factor $U(\mu_j,\mu)$ is evaluated numerically.

\subsubsection{Narrow resolution}
As for the photon jet function discussed in Section \ref{app:photon_jetFunc}, the unobserved jet function in the narrow resolution case receives contributions from the index combinations $J^{33}(p^2)$,  $J^{34}(p^2)$,  $J^{43}(p^2)$ and  $J^{44}(p^2)$. The function $J^{33}(p^2)$ was already computed in \cite{Beneke:2018ssm} and a detailed discussion of its derivation is presented in \cite{Beneke:2019vhz}. The result is a complicated function of the masses of the SM particles, the invariant mass of the jet and of the virtuality and rapidity scales 
$\mu$ and $\nu$, respectively. The computation of $J^{34}(p^2)$,  $J^{43}(p^2)$ and  $J^{44}(p^2)$ is similar to the case of the photon jet function. One has to be careful which functions receive contributions from Wilson line diagrams and whether the gauge boson crossing the cut originates from a $W^3$ or a $B$ boson, in order to determine the correct prefactor using~\eqref{eq:rotMatrix}. Since $J^{34}(p^2) = J^{43}(p^2)$ we will content ourselves with presenting the results for $J^{34}(p^2)$ and $J^{44}(p^2)$. They read
\begin{align}
J^{34} (p^2, \mu, \nu) =& -\hat{s}_W(\mu) \hat{c}_W(\mu) \delta(p^2) + \hat{s}_W(\mu) \hat{c}_W(\mu) \delta(p^2 - m_Z^2) \nn \\
&+ J^{34}_{\text{Wilson}} (p^2, \mu, \nu) + J^{34}_{\text{se},\, f\neq t\, \text{only}}(p^2,\mu) + J^{34}_{\text{se},\, f\neq t\, \text{excluded}}(p^2,\mu)\,.
\end{align}
\begin{align}
J^{34}_{\text{Wilson}}&(p^2,\mu,\nu) =  \frac{\hat{s}_W(\mu)\hat{c}_W(\mu) \hat{g}^2_2(\mu)}{16 \pi^2} \bigg\{ \delta(p^2)\bigg[ - 8\ln\frac{m_W}{\mu} \ln\frac{2 m_\chi}{\nu}+ 4\ln\frac{m_W}{\mu}\bigg] \, \nonumber \\
&+ \frac{1}{p^2}\ \theta(p^2-4 m^2_W)\bigg[ 2 \beta + 4\ln\frac{1-\beta}{1+\beta} \bigg]\bigg\} \, \nonumber \\
&  - \frac{\hat{s}_W(\mu)\hat{c}_W(\mu) \hat{g}^2_2(\mu)}{16 \pi^2} \bigg\{ \delta(p^2-m^2_Z) \bigg[- 8\ln\frac{m_W}{\mu} \ln\frac{2 m_\chi}{\nu} + 4\ln\frac{m_W}{\mu}- 4 + 2\pi^2   \, \nonumber\\
&+ 2 \pi \bar{\beta}_Z  - (8 \pi+4 \bar{\beta}_Z)\arctan(\bar{\beta}_Z)  + 8 \arctan^2(\bar{\beta}_Z)\bigg] \, 
\nonumber \\
&+ \frac{1}{p^2-m^2_Z} \,\theta(p^2-4 m^2_W)
\bigg[ 2 \beta + 4\ln\frac{1-\beta}{1+\beta} \bigg]\bigg\} \, ,
\end{align}
\begin{align}
J^{34}_{\text{se},\, f\neq t\, \text{only}}(p^2,\mu) &= \frac{\hat{s}^2_W(\mu) \hat{g}^2_2(\mu)}{16 \pi^2} \bigg\{ -\hat{s}_W(\mu )\hat{c}_W(\mu)  \frac{80}{9} \bigg[-\delta(p^2) \frac{5}{3}  + \bigg[\frac{1}{p^2}\bigg]^{[\mu^2]}_*\bigg]\nonumber \\
& + (\hat{s}^2_W(\mu) - \hat{c}^2_W(\mu))  \,\bigg(\frac{10}{3}\frac{1}{\hat{s}_W(\mu)\hat{c}_W(\mu)} -\frac{80}{9}\frac{\hat{s}_W(\mu)}{\hat{c}_W(\mu)}  \bigg) \nonumber \\
&\times \bigg[\bigg[\frac{1}{p^2-m^2_Z}\bigg]_* - \delta(p^2-m^2_Z)\nonumber \bigg( \frac{5}{3}-\ln\frac{m^2_Z}{\mu^2}\bigg)\bigg] \nonumber \\
&+ \hat{s}_W(\mu)\hat{c}_W(\mu) \bigg(-\frac{20}{3}\frac{1}{\hat{c}^2_W(\mu)} +\frac{7}{2}\frac{1}{\hat{s}^2_W(\mu)\hat{c}^2_W(\mu)}+\frac{80}{9}\frac{\hat{s}^2_W(\mu)}{\hat{c}^2_W(\mu)}\bigg) \nonumber \\
&\times \bigg[\bigg[\frac{1}{(p^2-m^2_Z)^2}\bigg]_{**} p^2 -  \delta(p^2-m^2_Z) \bigg(\frac{2}{3}-\ln\frac{m^2_Z}{\mu^2}\bigg)\bigg]\bigg\} \, .
\end{align}
\begin{align}
&\hspace*{-1cm} 
J^{34}_{\text{se},\, f\neq t\, \text{excluded}}(p^2,\mu) = \nonumber \, \\
&\left(\hat{s}^2_W(\mu) - \hat{c}^2_W(\mu)\right) \bigg[\frac{\text{Re}\big[\Sigma^{\gamma Z}_T(0)\big]_{t,W}}{m^2_Z}\delta(p^2) - \frac{\text{Re}\big[\Sigma^{\gamma Z}_T(m^2_Z)\big]_{t,W}}{m^2_Z} \delta(p^2-m^2_Z)\bigg] \nonumber \\
+& \,\hat{s}_W(\mu)\hat{c}_W(\mu) \,\text{Re}\frac{\partial \Sigma^{\gamma \gamma}_T(p^2)_{t,W}}{\partial p^2}\bigg|_{p^2=0} \delta(p^2) 
 \nonumber \\
-&\, \hat{s}_W(\mu)\hat{c}_W(\mu) \,\text{Re}\frac{\partial \Sigma^{ZZ}_T(p^2)_{t,W,Z,H}}{\partial p^2}\bigg|_{p^2=m^2_Z}\delta(p^2-m^2_Z) \nonumber \\
+& \left(\hat{s}^2_W(\mu) - \hat{c}^2_W(\mu)\right) \bigg[-\frac{1}{m^2_Z}\frac{1}{p^2} \frac{\text{Im}\big[\Sigma^{\gamma Z}_T(p^2)\big]_{t,W}}{\pi} +\frac{1}{m^2_Z}\frac{1}{p^2-m^2_Z} \frac{\text{Im}\big[\Sigma^{\gamma Z}_T(p^2)\big]_{t,W}}{\pi}\bigg] \nonumber \\
-&\, \hat{s}_W(\mu)\hat{c}_W(\mu) \frac{1}{\big(p^2\big)^2} \frac{\text{Im} \big[\Sigma^{\gamma \gamma}_T(p^2)\big]_{t,W}}{\pi} + \hat{s}_W(\mu)\hat{c}_W(\mu) \frac{1}{\big(p^2-m^2_Z\big)^2} \frac{\text{Im} \big[\Sigma^{Z Z}_T(p^2)\big]_{t,W,Z,H}}{\pi} \nonumber \\[-0.5cm]
&
\end{align}
for $J^{34}(p^2)$ and
\begin{align}
J^{44} = \hat{c}_W^2 \delta(p^2) + \hat{s}_W^2 \delta(p^2 - m_Z^2) + J^{44}_{\text{se},\, f\neq t\, \text{only}}(p^2,\mu) + J^{44}_{\text{se},\, f\neq t\, \text{excluded}}(p^2,\mu) \,,
\end{align}
\begin{align}
J^{44}_{\text{se},\, f\neq t\, \text{only}}(p^2,\mu) &= \frac{\hat{c}^{2}_W(\mu) \hat{g}^2_1(\mu)}{16 \pi^2} \Bigg\{ \hat{c}^2_W(\mu )  \frac{80}{9} \bigg[-\delta(p^2) \frac{5}{3}  + \bigg[\frac{1}{p^2}\bigg]^{[\mu^2]}_*\bigg]\nonumber \\
& - 2  \,\bigg(\frac{10}{3} -\frac{80}{9} \hat{s}^2_W(\mu) \bigg)  
\bigg[\bigg[\frac{1}{p^2-m^2_Z}\bigg]_* - \delta(p^2-m^2_Z)\nonumber \bigg( \frac{5}{3}-\ln\frac{m^2_Z}{\mu^2}\bigg)\bigg] \nonumber \\
&+ \frac{\hat{s}_W^2(\mu)}{\hat{c}_W^2(\mu)} \bigg(-\frac{20}{3} +\frac{7}{2}\frac{1}{\hat{s}^2_W(\mu)}+\frac{80}{9}\hat{s}^2_W(\mu)\bigg) \nonumber \\
&\times \bigg[\bigg[\frac{1}{(p^2-m^2_Z)^2}\bigg]_{**} p^2 - \delta(p^2-m^2_Z) \bigg(\frac{2}{3}-\ln\frac{m^2_Z}{\mu^2}\bigg)\bigg]\Bigg\} \, .
\end{align}
\begin{align}
&\hspace*{-1cm} J^{44}_{\text{se},\, f\neq t\, \text{excluded}}(p^2,\mu) = \nonumber \, \\
&-2 \hat{s}_W(\mu) \hat{c}_W(\mu) \bigg[\frac{\text{Re}\big[\Sigma^{\gamma Z}_T(0)\big]_{t,W}}{m^2_Z}\delta(p^2) - \frac{\text{Re}\big[\Sigma^{\gamma Z}_T(m^2_Z)\big]_{t,W}}{m^2_Z} \delta(p^2-m^2_Z)\bigg] \nonumber \\
-& \hat{c}^2_W(\mu) \text{Re}\frac{\partial \Sigma^{\gamma \gamma}_T(p^2)_{t,W}}{\partial p^2}\bigg|_{p^2=0} \delta(p^2) - \hat{s}^2_W(\mu) \text{Re}\frac{\partial \Sigma^{ZZ}_T(p^2)_{t,W,Z,H}}{\partial p^2}\bigg|_{p^2=m^2_Z}\delta(p^2-m^2_Z) \nonumber \\
-& 2 \hat{s}_W(\mu) \hat{c}_W(\mu) \bigg[-\frac{1}{m^2_Z}\frac{1}{p^2} \frac{\text{Im}\big[\Sigma^{\gamma Z}_T(p^2)\big]_{t,W}}{\pi} +\frac{1}{m^2_Z}\frac{1}{p^2-m^2_Z} \frac{\text{Im}\big[\Sigma^{\gamma Z}_T(p^2)\big]_{t,W}}{\pi}\bigg] \nonumber \\
& + \hat{c}^2_W(\mu) \frac{1}{\big(p^2\big)^2} \frac{\text{Im} \big[\Sigma^{\gamma \gamma}_T(p^2)\big]_{t,W}}{\pi} + \hat{s}^2_W(\mu) \frac{1}{\big(p^2-m^2_Z\big)^2} \frac{\text{Im} \big[\Sigma^{Z Z}_T(p^2)\big]_{t,W,Z,H}}{\pi} 
\end{align}
for $J^{44}(p^2)$.

\subsection{Soft function}
\label{app:soft_func}

\subsubsection{Intermediate resolution}
In this appendix, we collect the soft functions appearing in the factorization theorem for the intermediate resolution case~\eqref{eq:logfact} and discuss their resummation. We only give the non-vanishing soft-function coefficients. They read 
\begin{align}
W_{IJ, \,33}^{{\rm SU(2)},\,11}\left(\omega,\mu,\nu\right) &= 
\delta(\omega) + \frac{\hat{g}_2^2(\mu)}{16 \pi^2} \nn \\
&\times \,\bigg\{ \delta(\omega) \left[-16 \ln \frac{m_W}{\mu} \ln\frac{m_W}{\nu}\right] + \left[\frac{1}{\omega}\right]_*^{\left[m_W\right]} \left[-16 \ln\frac{m_W}{\mu}\right] \bigg\} \,,\quad\\[0.1cm]
W_{IJ, \,34}^{{\rm SU(2)},\,14}\left(\omega,\mu,\nu\right) &= 
\frac{n_{IJ}^{14}}{2}\delta(\omega) + 
\frac{n_{IJ}^{14}}{2}\frac{\hat{g}_2^2(\mu)}{16 \pi^2} 
\bigg\{ \delta(\omega) \left[ \frac{\pi^2}{6}
-(4+8i\pi)\ln\frac{m_W}{\mu} \right. \nn \\ 
&\left. - 8\ln\frac{m_W}{\mu}\ln\frac{m_W}{\nu} -4\ln^2\frac{m_W}{\mu} \right]  \nn \\
& -8\left[\frac{1}{\omega}\right]_*^{\left[m_W\right]} \left[ \ln\left(\frac{m_W^2 + \omega^2}{m_W^2}\right)  + \ln\frac{m_W^2}{\mu^2}\right] \bigg\} \,,\\[0.1cm]
W_{IJ, \,43}^{{\rm SU(2)},\,41}\left(\omega,\mu,\nu\right) &= W^{{\rm SU(2)},\,14 \, *}_{JI,34}\left(\omega,\mu,\nu\right)\,, \\[0.1cm]
W_{IJ, \,44}^{{\rm SU(2)},\,44}\left(\omega,\mu\right) &=  \frac{n_{IJ}^{44}}{4}\delta(\omega) + \frac{n_{IJ}^{44}}{4}\frac{\hat{g}_2^2(\mu)}{16 \pi^2} \left\{ \delta(\omega) \left[ \frac{\pi^2}{3} -8\ln\frac{m_W}{\mu}-8\ln^2\frac{m_W}{\mu} \right]\right. \nn \\
&\left. +8\left[\frac{1}{\omega}\right]_*^{\left[m_W\right]} \left[- \ln\left(\frac{m_W^2 + \omega^2}{m_W^2}\right) - \ln\frac{m_W^2}{\mu^2} \right] -8\frac{\omega}{m_W^2 +\omega^2} \right\},\\[0.1cm]
W_{IJ, \,33}^{{\rm U(1)},\,44}\left(\omega,\mu,\nu\right) &=\, \frac{n_{IJ}^{44}}{4}\delta(\omega) + \frac{n_{IJ}^{44}}{4}\frac{\hat{g}_2^2(\mu)}{16 \pi^2}  \nn \\ 
&\times \left\{ \delta(\omega) \left[ -\frac{\pi^2}{3} -8 \ln \frac{m_W}{\mu} -16 \ln \frac{m_W}{\mu} \ln\frac{m_W}{\nu} +8\ln^2\frac{m_W}{\mu} \right] \right\},\\[0.1cm]
W_{IJ, \,34}^{{\rm U(1)},\,46}\left(\omega,\mu,\nu\right) &=\, \frac{n_{IJ}^{46}}{2}\delta(\omega) + \frac{n_{IJ}^{46}}{2}\frac{\hat{g}_2^2(\mu)}{16 \pi^2}  \nn \\ 
&\times \left\{ \delta(\omega) \left[ -\frac{\pi^2}{6} -4 \ln \frac{m_W}{\mu} -8 \ln \frac{m_W}{\mu} \ln\frac{m_W}{\nu} +4\ln^2\frac{m_W}{\mu} \right]\right\},\\[0.1cm]
W_{IJ, \,43}^{{\rm U(1)},\,64}\left(\omega,\mu,\nu\right) &= W_{JI, \,34}^{{\rm U(1)},\,46 \, *}\left(\omega,\mu,\nu\right)\\[0.1cm]
W_{IJ, \,44}^{{\rm U(1)},\,66}\left(\omega\right) &=\, \delta\left(\omega\right) \,,
\end{align}
where we introduced
\begin{align}
n_{IJ}^{ij} &= (-1)^{\delta_{I (00)} \delta_{i4}} (-1)^{\delta_{J (00)} \delta_{j 4}}\quad {\rm with} \quad (00) = (11) \ {\rm or} \ (22),
\end{align}
to allow for a more compact notation. The resummation of the soft function is conceptually analogous to the resummation of the soft function in the case of wino DM \cite{Beneke:2019vhz}, to which we refer to for a detailed discussion. The Laplace transform of the soft function 
and its inverse are defined as
\begin{align}
\label{eq:softLapTrans}
w(s) &= \mathcal{L}\,\{W(\omega)\} = 
\int_0^{\infty} d \omega \, e^{-\omega s} \,W(\omega) \,, 
\\
W(\omega) &= \mathcal{L}^{-1}\left\{w(s)\right\} = 
\frac{1}{2 \pi i} \int^{c+i \infty}_{c-i \infty} 
d s\,  e^{s \omega} \,w(s) \, ,
\end{align}
where $s=1/(e^{\gamma_E}\kappa)$. The Laplace transforms required are
\begin{align}
\mathcal{L}\left\{\delta ( \omega )\right\}&=1 \, , \nonumber \\
\mathcal{L}\left\{\left[\frac{1}{\omega}\right]_*^{\left[m_W\right]}\right\}&= \ln\left(\frac{\kappa}{m_W}\right) \, , \nonumber \\
\mathcal{L}\left\{\frac{1}{\omega}\ln \left(\frac{m_W^2+\omega^2}{m_W^2}\right)\right\}&= 
\text{si}^2\left(m_W s\right) + \text{ci}^2\left(m_W s\right)\equiv\tilde{G}(s) \, , \nonumber \\
\mathcal{L}\left\{\frac{\omega}{m_W^2 + \omega^2}\right\} &= \cos(m_W s) \text{ci}(m_W s) -\sin(m_W s) \text{si}(m_W s)\equiv \tilde{Q}(s) \, ,
\label{eq:LaplaceTransforms}
\end{align}
where the functions si, ci are defined as
\begin{align}
\text{si}(x) \equiv -\int_{x}^{\infty}dt\, \frac{\sin(t)}{t} \quad \text{and}\quad \text{ci}(x) \equiv -\int_{x}^{\infty} dt\, \frac{\cos(t)}{t}\, .
\end{align}
For convenience, we introduce a vector notation for the 
soft functions as follows:
\begin{align}
\vec{w}_{IJ}^{\, \rm SU(2)} =& \left(w_{IJ, 33}^{\text{SU(2)},11}, w_{IJ,34}^{\text{SU(2)},14}, w_{IJ,43}^{\text{SU(2)},41}, w_{IJ,44}^{\text{SU(2)},44}\right)^T \,, \nn \\
\vec{w}_{IJ}^{\, \rm U(1)} =& \left(w_{IJ,33}^{\text{U(1)},44}, w_{IJ,34}^{\text{U(1)},46}, w_{IJ,43}^{\text{U(1)},64}, w_{IJ,44}^{\text{U(1)},66}\right)^T \,.
\end{align}
The rapidity RG equations for the soft functions take the form
\begin{align}
\label{eq:softRRG}
\frac{d}{d\ln\nu}\vec{w}_{IJ}^{\, \text{SU(2)}}\left(s,\mu,\nu\right) = \mathbf{\Gamma}_W^{\text{SU(2)},\nu} \vec{w}_{IJ}^{\,\text{SU(2)}}\left(s,\mu,\nu\right) \,, \nn \\
\frac{d}{d\ln\nu}\vec{w}_{IJ}^{\,\text{U(1)}}\left(s,\mu,\nu\right) = \mathbf{\Gamma}_W^{\text{U(1)},\nu} \vec{w}_{IJ}^{\,\text{U(1)}}\left(s,\mu,\nu\right)
\end{align}
with one-loop rapidity anomalous dimensions given by
\begin{align}
\mathbf{\Gamma}_W^{\text{SU(2)},\nu} = \mathbf{\Gamma}_W^{\text{U(1)},\nu} = \frac{\hat{\alpha}_2(\mu)}{4\pi}2\gamma_{\rm cusp}^{(0)}\ln\frac{m_W}{\mu} \,\text{diag}\left[2,1,1,0\right] \,.
\label{eq:softRRG1}
\end{align}
The RG equations are given by
\begin{align}
\label{eq:softRG}
\frac{d}{d\ln\mu}\vec{w}_{IJ}^{\,\text{SU(2)}}\left(s,\mu,\nu\right) = \mathbf{\Gamma}_W^{\text{SU(2)},\mu} \vec{w}_{IJ}^{\,\text{SU(2)}}\left(s,\mu,\nu\right) \,, \nn \\
\frac{d}{d\ln\mu}\vec{w}_{IJ}^{\,\text{U(1)}}\left(s,\mu,\nu\right) = \mathbf{\Gamma}_W^{\text{U(1)},\mu} \vec{w}_{IJ}^{\,\text{U(1)}}\left(s,\mu,\nu\right) \,,
\end{align}
with the anomalous dimensions
\begin{eqnarray}
\label{eq:GammaADMs}
\mathbf{\Gamma}_W^{\text{SU(2)},\mu} &=& 4\gamma_{\rm cusp} \ln\frac{\kappa}{\nu} \,\mathbf{1}_4
+ 2\gamma_{\rm cusp}\left(\ln\frac{m_W}{\mu} - \ln\frac{m_W}{\nu}\right) \text{diag}\left[0,1,1,2\right] \nn \\
&&+ \,\text{diag}\left[0, \gamma_W, \gamma_W^*, 2\text{Re}\,\gamma_W\right] \,, \nn \\
\mathbf{\Gamma}_W^{\text{U(1)},\mu} &=& 2\gamma_{\rm cusp}\left(\ln\frac{m_W}{\nu} - \ln\frac{m_W}{\mu}\right) \,\text{diag}\left[2,1,1,0\right] \nn \\
&&+\,
\text{diag}\left[2\text{Re}\, \gamma_W, \text{Re}\, \gamma_W, \text{Re}\, \gamma_W^*, 0\right]\,.
\end{eqnarray}
At the one-loop order, which is sufficient for NLL' resummation, the anomalous dimension $\gamma_W$ evaluates to
\begin{align}
\gamma_W^{(0)} = 4 + 8i\pi\,.
\end{align}
Since the anomalous dimensions appearing in the rapidity RG~\eqref{eq:softRRG} and RG~\eqref{eq:softRG} equations are diagonal, their solutions are straightforward. Note that the rapidity anomalous dimensions~\eqref{eq:softRRG1} are independent of the scale $\nu$, which allows us to find the analytic solutions
\begin{align}
\vec{w}_{IJ}^{\,\rm SU(2)}\left(s,\mu,\nu\right) =& \,\exp\left[\mathbf{\Gamma}_W^{\text{SU(2)},\nu}\ln\frac{\nu}{\nu_s}\right] \vec{w}_{IJ}^{\,\rm SU(2)}\left(s,\mu,\nu_s\right) \,, \nn \\
\vec{w}_{IJ}^{\,\rm U(1)}\left(s,\mu,\nu\right) =& \,\exp\left[\mathbf{\Gamma}_W^{\text{U(1)},\nu}\ln\frac{\nu}{\nu_s}\right] \vec{w}_{IJ}^{\,\rm U(1)}\left(s,\mu,\nu_s\right) \,,
\label{eq:RRGevolFacts}
\end{align}
while the solution of the RG equations is given by
\begin{eqnarray}
\vec{w}_{IJ}^{\,\rm SU(2)}\left(s,\mu,\nu\right) &=& \mathbf{U}_W^{\text{SU(2)}, \mu}\left(\mu,\mu_s,\nu\right) \vec{w}_{IJ}^{\,\rm SU(2)}\left(s,\mu_s,\partial_\eta\right) \, \left(\frac{\kappa}{\nu}\right)^{\eta} \,, \nn \\
\vec{w}_{IJ}^{\,\rm U(1)}\left(s,\mu,\nu\right) &=& \mathbf{U}_W^{\text{U(1)}, \mu}\left(\mu,\mu_s,\nu\right) \vec{w}_{IJ}^{\,\rm U(1)}\left(s,\mu_s,\nu\right) \,.
\label{eq:evolvedLapW}
\end{eqnarray}
In~\eqref{eq:evolvedLapW} we introduced the variable $\eta$  
defined as 
\begin{align}
\eta = 4 \int_{\ln \mu_s}^{\ln \mu} d\ln\mu' \,\gamma_{\rm cusp} \left(\hat{\alpha}_2\left(\mu'\right)\right)\,.
\end{align}
The evolution matrices $\mathbf{U}_W^{\text{SU(2)}, \mu}$ and $\mathbf{U}_W^{\text{U(1)}, \mu}$ are
\begin{align}
\mathbf{U}_W^{\text{SU(2)}, \mu}\left(\mu,\mu_s,\nu\right) =& \,\exp\left[\int_{\ln \mu_s}^{\ln \mu} d\ln\mu'\, \mathbf{\Gamma}_{W,\gamma_W}^{\text{SU(2)},\mu}\right]\,, \nn \\
\mathbf{U}_W^{\text{U(1)}, \mu}\left(\mu,\mu_s,\nu\right) =& \,\exp\left[\int_{\ln \mu_s}^{\ln \mu} d\ln\mu'\, \mathbf{\Gamma}_W^{\text{U(1)},\mu}\right]\,,
\end{align}
where $ \mathbf{\Gamma}_{W,\gamma_W}^{\text{SU(2)},\mu}$ is 
$\mathbf{\Gamma}_{W}^{\text{SU(2)},\mu}$ given in 
(\ref{eq:GammaADMs}) with the $\kappa$-dependent cusp term removed.
Now that we have computed the evolution factors, we perform the 
inverse Laplace transforms in order to return to momentum space. For 
$\vec{w}_{IJ}^{\,\rm U(1)}$ the inverse transform is trivial since it contains no dependence on $\kappa$. For the inverse Laplace transform, we define
\begin{align}
\hat{\vec{W}}_{IJ}^{\rm SU(2)}\left(\omega,\mu,\nu\right) = \mathcal{L}^{-1} \left[\vec{w}_{IJ}^{\,\rm SU(2)}\left(s,\mu,\partial_\eta\right) \, \left(\frac{\kappa}{\nu}\right)^{\eta}\right] \,,
\end{align}
and make use of the relations
\begin{align}
&\mathcal{L}^{-1}\left[1\right] = \delta(\omega) \, , \nn \\
&\mathcal{L}^{-1}\left[\left(\frac{\kappa}{\nu}\right)^\eta\right]  
= \frac{e^{-\gamma_E \eta} }{\Gamma(\eta)}\left(\frac{\omega}{\nu}\right)^\eta\frac{1}{\omega} \, , \nn \\
&F(\omega) \equiv \mathcal{L}^{-1}\left[ 
\left(\frac{\kappa}{\nu}\right)^\eta \tilde{G}\big(e^{-\gamma_E}/\kappa
\big)\right] 
\nonumber\\
&\hspace*{1cm} = 
\left(\frac{e^{-\gamma_E}}{\nu}\right)^{\eta}\frac{\omega^{1+\eta}}{m_W^2 \Gamma(2+\eta)} \,
{}_4F_3\left(1,1,1,\frac{3}{2};1+\frac{\eta}{2},\frac{3}{2}+\frac{\eta}{2},2;-\frac{\omega^2}{m_W^2}\right),\nn \\
&P(\omega) \equiv \mathcal{L}^{-1}\left[
\left(\frac{\kappa}{\nu}\right)^\eta \tilde{Q}\big(e^{-\gamma_E}/\kappa\big)\right]
\nonumber\\
&\hspace*{1cm} =\left(\frac{e^{-\gamma_E}}{\nu}\right)^\eta \frac{\omega^{1+\eta}}{m_W^2 \Gamma(2+\eta)}\, {}_{3}F_2\left(1,1,\frac{3}{2};1+\frac{\eta}{2},\frac{3}{2}+\frac{\eta}{2};-\frac{\omega^2}{m_W^2}\right) \label{eq:invTrans} .
\end{align}
Finally, we find that the virtuality-resummed soft function in momentum space takes the form
\begin{eqnarray}
\vec{W}_{IJ}^{\rm SU(2)}\left(\omega,\mu,\nu\right) &=& \mathbf{U}_W^{\text{SU(2)}, \mu}\left(\mu,\mu_s,\nu\right) \,\hat{\vec{W}}_{IJ}^{\rm SU(2)}\left(\omega,\mu_s,\nu\right) \,, \nn \\
\vec{W}_{IJ}^{\rm U(1)}\left(\omega,\mu,\nu\right) &=& \mathbf{U}_W^{\text{U(1)}, \mu}\left(\mu,\mu_s,\nu\right) \,\vec{W}_{IJ}^{\rm U(1)}\left(\omega,\mu_s,\nu\right) \,.
\label{eq:resummedSoft}
\end{eqnarray}
As in \cite{Beneke:2019vhz}, we did not include the rapidity evolution factor in~\eqref{eq:resummedSoft}, since we evolve the photon jet function in $\nu$ from $\nu_h$ to $\nu_s$, which makes the soft function rapidity evolution factors unity (see~\eqref{eq:RRGevolFacts}).

\subsubsection{Narrow resolution}

In this appendix, we collect the soft functions appearing in the factorization theorem for the narrow resolution case \eqref{eq:nrwlogfact}. The definition and computation of these terms is analogous to the wino DM case and we refer the reader to \cite{Beneke:2018ssm} for more details. The expressions take the following form:
\begin{eqnarray}
D^1_{(00),33} &=& 1+\frac{\hat{g}_2^2(\mu)}{16\pi^2} \left[-\frac{\pi^2}{3} +8i\pi\ln\frac{m_W}{\mu} -16\ln\frac{m_W}{\mu}\ln\frac{m_W}{\nu} +8\ln^2\frac{m_W}{\mu} \right] \,, \nn \\
D^1_{(+-),33} &=& D^1_{(00),33} \,,\nn \\[0.1cm]
D^4_{(00),34} &=& -\frac{1}{2}-\frac{1}{2}\frac{\hat{g}_2^2(\mu)}{16\pi^2} \left[ -\frac{\pi^2}{6} -4\ln\frac{m_W}{\mu} -8\ln\frac{m_W}{\mu}\ln\frac{m_W}{\nu} +4\ln^2\frac{m_W}{\mu} \right]  \,, \nn \\
D^4_{(+-),34} &=& -D^4_{(00),34}  \,,\nn\\[0.1cm]
D^4_{(00),43} &=& D^4_{(00),34}  \,, \quad
D^4_{(+-),43} = D^4_{(+-),34}  \,,\nn \\[0.1cm]
D^6_{(00),44} &=& 1  \,, \quad
D^6_{(+-),44} = 1  \,.
\end{eqnarray}


\bibliography{mybib}

%
%


\end{document}